\documentclass[fleqn,usenatbib]{mnras}

\usepackage[T1]{fontenc}

\usepackage{newtxtext,newtxmath}

\usepackage{graphicx}	
\usepackage{amsmath}	
\usepackage{amssymb}	

\usepackage{xspace}

\usepackage{csquotes}

\newcommand{\fref}[1]{Fig.~\ref{#1}\xspace}
\newcommand{\eref}[1]{eq.~\eqref{#1}\xspace}
\newcommand{\sref}[1]{Section~\ref{#1}\xspace}
\newcommand{\tref}[1]{Table~\ref{#1}\xspace}
\newcommand{\aref}[1]{Appendix~\ref{#1}\xspace}

\newcommand{\papI}{Paper I\xspace}
\newcommand{\pp}{J17\xspace}

\newcommand{\papRGt}{Silva Aguirre et al. (submitted)\xspace}
\newcommand{\papRGp}{Silva Aguirre et al., submitted\xspace}
\newcommand{\papRGiip}{Christensen-Dalsgaard et al., submitted\xspace}

\newcommand{\msun}{\ensuremath{\, \mathrm{M}_{\odot}}\xspace}
\newcommand{\rsun}{\ensuremath{\, \mathrm{R}_{\odot}}\xspace}
\newcommand{\tsun}{\ensuremath{\, \mathrm{T}_{\mathrm{eff} , \odot}}\xspace}
\newcommand{\kel}{\ensuremath{\, \mathrm{K}}\xspace}
\newcommand{\dex}{\ensuremath{\, \mathrm{dex}}\xspace}
\newcommand{\mm}{\ensuremath{\, \mathrm{Mm}}\xspace}
\newcommand{\muhz}{\ensuremath{\, \mu\mathrm{Hz}}\xspace}

\newcommand{\gar}{\textsc{garstec}\xspace}

\newcommand{\adi}{\textsc{adipls}\xspace}
\newcommand{\bas}{\textsc{basta}\xspace}
\newcommand{\leg}{\textsc{legacy}\xspace}
\newcommand{\kep}{\textsl{Kepler}\xspace}
\newcommand{\corot}{\textsl{CoRoT}\xspace}
\newcommand{\staggrid}{Stagger grid\xspace}
\newcommand{\rtgrid}{Trampedach grid\xspace}

\newcommand{\ttau}{\ensuremath{T(\tau)}\xspace}
\newcommand{\ttaurel}{\ttau relation\xspace}
\newcommand{\feh}{\ensuremath{[\mathrm{Fe}/\mathrm{H}]}\xspace}
\newcommand{\td}{\ensuremath{\langle \mathrm{3D} \rangle}\xspace}
\newcommand{\tde}{\td-envelope\xspace}
\newcommand{\tdes}{\tde{}s\xspace}
\newcommand{\mlt}{\ensuremath{\alpha_{\rm\scriptscriptstyle{MLT}}}\xspace}
\newcommand{\logg}{\ensuremath{\log g}\xspace}
\newcommand{\teff}{\ensuremath{T_{\mathrm{eff}}}\xspace}
\newcommand{\pgas}{\ensuremath{P_{\mathrm{gas}}}\xspace}
\newcommand{\rma}{\ensuremath{r_{\mathrm{m}}}\xspace}
\newcommand{\dma}{\ensuremath{d_{\mathrm{m}}}\xspace}
\newcommand{\tma}{\ensuremath{T_{\mathrm{m}}}\xspace}

\newcommand{\adgrad}{\ensuremath{\nabla_{\mathrm{ad}}}\xspace}
\newcommand{\gam}{\ensuremath{\Gamma_{1}}\xspace}
\newcommand{\gamm}{\ensuremath{\Gamma_{1}^{\mathrm{1D}}}\xspace}
\newcommand{\gammm}{\ensuremath{\Gamma_{1}^{\mathrm{3D}}}\xspace}
\newcommand{\ys}{\ensuremath{Y_{\mathrm{s}}}\xspace}
\newcommand{\yi}{\ensuremath{Y_{\mathrm{i}}}\xspace}
\newcommand{\zi}{\ensuremath{Z_{\mathrm{i}}}\xspace}
\newcommand{\rcz}{\ensuremath{r_{\mathrm{cz}}}\xspace}
\newcommand{\otf}{on-the-fly\xspace}
\newcommand{\numax}{\ensuremath{\nu_{\mathrm{max}}}\xspace}
\newcommand{\mean}[1]{\ensuremath{\langle #1 \rangle}\xspace}
\newcommand{\dnu}{\ensuremath{\Delta\nu}\xspace}
\newcommand{\km}{\ensuremath{K_{\mathrm{m}}}\xspace}

\title[Coupling 1D evolution with 3D simulations]%
{Coupling 1D stellar evolution with 3D-hydrodynamical simulations on-the-fly II:
  Stellar Evolution and Asteroseismic Applications}

\author[J. R. Mosumgaard et al.]{%
  Jakob R{\o}rsted Mosumgaard$^{1,2}$\thanks{E-mail: jakob@phys.au.dk},
  Andreas Christ S{\o}lvsten J{\o}rgensen,$^{2}$\thanks{E-mail: acsj@mpa-garching.mpg.de},
  Achim Weiss$^{2}$,
  \and
  V\'{i}ctor Silva Aguirre$^{1}$
  and
  J{\o}rgen Christensen-Dalsgaard$^{1,3}$
  \\
  $^{1}$ Stellar Astrophysics Centre (SAC), Department of Physics and
  Astronomy, Aarhus University, Ny Munkegade 120, DK-8000 Aarhus C, Denmark\\
  $^{2}$ Max-Planck-Institut f\"{u}r Astrophysik, Karl-Schwarzschild-Str. 1,
  D-85748 Garching, Germany\\
  $^{3}$ Kavli Institute for Theoretical Physics, University of California Santa Barbara, CA 93106-4030\\
}

\date{Accepted 2019 October 21. Received 2019 October 13; in original form 2019 July 7}

\pubyear{2019}

\begin{document}
\label{firstpage}
\pagerange{\pageref{firstpage}--\pageref{lastpage}}
\maketitle

\begin{abstract}
  Models of stellar structure and evolution are an indispensable tool in
  astrophysics, yet they are known to incorrectly reproduce the outer convective
  layers of stars. In the first paper of this series, we presented a novel
  procedure to include the mean structure of 3D hydrodynamical simulations \otf
  in stellar models, and found it to significantly improve the outer
  stratification and oscillation frequencies of a standard solar model. In the
  present work, we extend the analysis of the method; specifically how the
  transition point between envelope and interior affects the models. We confirm
  the versatility of our method by successfully repeating the entire procedure
  for a different grid of 3D hydro-simulations. Furthermore, the applicability
  of the procedure was investigated across the HR diagram and an accuracy
  comparable to the solar case was found. Moreover, we explored the implications
  on stellar evolution and find that the red-giant branch is shifted about
  $40\kel$ to higher effective temperatures. Finally, we present for the first
  time an asteroseismic analysis based on stellar models fully utilising the
  stratification of 3D simulations \otf. These new models significantly reduce
  the asteroseismic surface term for the two selected stars in the \kep field.
  We extend the analysis to red giants and characterise the shape of the surface
  effect in this regime. Lastly, we stress that the interpolation required by
  our method would benefit from new 3D simulations, resulting in a finer
  sampling of the grid.
\end{abstract}

\begin{keywords}
  asteroseismology -- stars: atmospheres -- stars:evolution -- stars: interiors
  -- stars: oscillations -- stars: solar-type
\end{keywords}



\section{Introduction}
\label{sec:intro}

An essential part of stellar structure calculations is the description of
convection: the hydrodynamical phenomenon when energy cannot be transported
stably by radiation alone. One of the most successful approaches from an
astrophysical point of view is the mixing-length theory (MLT), which is an
approximative, parametric description of convection. The basic principle is to
model convection as rising and falling elements moving a certain distance in a
stable background medium before dissolving. As described by e.g.
\citet{gough76a}, the mixing-length approach to convection was developed in the
early 20th century by several independent authors. The MLT was applied in an
astronomical context by \citet{biermann32a}, and introduced to stellar modelling
in a generalised form by B\"{o}hm-Vitense~\citep{Vitense1953,Bohm-Vitense1958},
whose formalism has since been the most common one.

The advantage of utilizing such parametrizations is that the nuclear reactions
-- rather than the convection motion -- set the time scale for the evolution
calculations. A drawback of the MLT approach is that it involves free
parameters, several of which are specified by the employed MLT flavour
\citep[e.g.][]{Bohm-Vitense1958,cox68a,kippenhahn12a}. However, the most
important one -- the so-called mixing-length parameter \mlt{} -- must be
determined from a calibration, which is typically performed to reproduce the Sun
\citep[e.g.][]{gough76a,Pedersen1990,Salaris2008}.

The MLT description of convection is known to be inadequate
\citep[e.g.,][]{trampedach10a} and results in incorrectly modelled outer layers
in low-mass stars with convective envelopes \citep[e.g.,][]{Rosenthal1999}. This
gives rise to a systematic offset in the predicted oscillation frequencies,
which was observed in the Sun, and confirmed by helioseismology to stem from
near-surface deficiencies in the stellar models
\citep{Brown1984,christensen-dalsgaard84a,Christensen-Dalsgaard1988}. Today, it
is commonly referred to as the (asteroseismic) \emph{surface effect}, and it is
crucial to take it into consideration when using oscillation frequencies to
determine stellar parameters. In the context of helioseismic inversions, the
surface term is typically suppressed by introducing a constraint based on a
series of polynomials \citep[e.g. the review by][and references
therein]{christensen-dalsgaard02a}. In asteroseismology, a correction in the
form of an empirical prescription is commonly used; two of the most popular ones
are the power law by \citet{Kjeldsen2008}, and the prescription by
\citet{Ball2014}.

To improve the modelling of the outer structure, an option is to utilise 2D or
3D radiation-coupled hydrodynamics simulations of stellar surface convection.
These simulations require no parametric theory; the convection is generated from
fundamental physics by treating the interaction between radiation and matter.
Such 3D simulations are inherently more realistic and have altered our
understading of solar granulation and stellar convection (e.g.
\citet{Stein1989,Stein1998}; \citealt{Nordlund2009}).

The first attempts to include information from such simulations to adjust models
of the present Sun were made by \citet{Schlattl1997} and \citet{Rosenthal1999}.
Recently, several authors
\citep{Piau2014,Sonoi2015,Ball2016,Magic2016,Trampedach2017} have produced
stellar models, where the outer layers of the 1D structure are substituted by
the mean stratification of 3D simulations. This procedure is commonly referred
to as patching, and requires a high degree of physical consistency between the
two model parts. It also requires a careful fit of the 1D model to the 3D
counterpart to ensure that the fundamental stellar surface parameters of the two
match. Due to the high computational cost of hydrodynamical simulations, these
methods are based on pre-calculated 3D atmospheres/envelopes.

A major limitation of these patching procedures is that they have limited
applicability: They can only be used to analyse a star with parameters exactly
matching those of a computed simulation. In order to circumvent this,
\citet{Joergensen2017} -- hereafter \pp -- established a new method to
interpolate between 3D simulations in atmospheric parameters (effective
temperature, \teff, and surface gravity, \logg). The scheme was able to reliably
reproduce the mean structure of 3D envelopes from two existing sets of
simulations: The \staggrid \citep{Magic2013} and the grid from
\citet{Trampedach2013}. Currently, only interpolation between simulations at
solar metallicity -- as defined in the respective 3D-simulation grids -- are
supported. \pp was able to construct patched models for stars not matching any
existing 3D simulations -- including several stars observed by the \kep space
mission \citep{Borucki2010x,gilliland10a}.

An inherent drawback of the patching methods -- which is not remedied by the
\pp-interpolation -- is that they do not take information from 3D simulations
into account \emph{during} the entire evolution. Only at the final model the
outer layers are substituted; the evolution prior to this point is calculated
using a stellar evolution code using a standard \mlt-prescription and analytical
boundary condition.

In order to overcome this shortcoming, \citet{Trampedach2014a,Trampedach2014b}
took another path to including 3D information in stellar models. Utilizing the
hydro-grid from \citet{Trampedach2013}, they distilled each simulation into a
stratification of temperature $T$ as a function of optical depth $\tau$ -- a
so-called \ttaurel{}, which can be used as a boundary condition in a stellar
model -- capable of reproducing the 3D photospheric transition. They also
performed a corresponding calibration of \mlt, which can be used by stellar
evolution codes throughout the evolution (for stars inside the simulation grid).
How stellar evolution is affected by employing this parametrisation was
initially investigated by \citet{Salaris2015} and \citet{Mosumgaard2017x}.
Recently, \citet{Mosumgaard2018} published details on how to implement the
results in a stellar evolution code, and presented an in-depth analysis of the
structural and asteroseismic impact, which turns out to be quite limited.

In the first paper of this series, \citet{Joergensen2018} -- hereafter \papI{}
-- presented a novel method for using averaged 3D-envelopes \otf for stellar
evolution. The \tdes are both used as boundary conditions to determine the
interior structure and are appended in each iteration of the calculation -- thus
omitting the need for post-evolutionary patching. The analysis showed that our
new method leads to a calibrated solar model, whose outer layers closely mimic
the structure of the underlying 3D simulations. Moreover, this improvement of
the stellar structure is shown to partly eliminate the structural contribution
to the surface effect.

In the present work, we extend the investigation of using \tde \otf for stellar
evolution. We elaborate upon aspects not treated in \papI as well as extend the
analysis to stars of different parameters. We defer interpolation in metallicity
on-the-fly to a later paper. A scheme for interpolation in metallicity was
recently presented by \citet{Joergensen2019} but has not yet been implemented
into a stellar evolution code. In the current paper we neglect turbulent
pressure; however, an implementation of this was recently presented by
\cite{JoergensenWeiss2019}.

The paper is organised as follows. In the next section, the method is briefly
summarised, and in \sref{sec:sun} we investigate further aspects of the solar
model. We apply our procedure to a different grid of 3D simulations in
\sref{sec:sun_tramp}. The impact on stellar evolution and structure models is
examined in \sref{sec:evol}. In \sref{sec:kepler}, we analyse the asteroseismic
impact of our new models and extend the discussion to red giants in
\sref{sec:rgb}. Our concluding remarks are found in \sref{sec:conclusions}.

\section{Method}
\label{sec:method}

In this section, we briefly summarize key aspects of our method. For further
details on the implementation, we refer to \papI.

Our calculations are made with the Garching Stellar Evolution Code
\citep[\gar,][]{Weiss2008} combined with a grid of 3D hydro-simulations of
stellar surface convection. In this paper -- except for in \sref{sec:sun_tramp}
-- we use simulations from the \staggrid by \citet{Magic2013} at solar
metallicity.

We have employed the interpolation scheme from \pp -- using the current \teff
and \logg of the star -- to determine an interpolated mean stratification. The
fundamental quantity of the procedure is the gas pressure as a function of
temperature $\pgas(T)$ extracted from the full 3D simulations. Note that the
shallowest 3D simulation in the grid dictates the highest possible pressure in
the interpolated envelope (see \pp). \footnote{When determining the
  interpolation range, it is important to exclude the un-physical border regions
  that strongly reflect the chosen lower boundary conditions. We have excluded
  these zones based on the superadiabatic index: $\nabla-\adgrad$, where $\nabla
  = \mathrm{d} \ln T/ \mathrm{d} \ln P$. Rather than approaching adiabatic
  conditions, this quantity increases with depth in the un-physical regime.}

The obtained \tde is used to provide the outer boundary conditions for solving
the stellar structure equations in the stellar model. In our implementation, the
boundary conditions are established deep within the superadiabatic layer -- we
refer to the corresponding point in the model as the \emph{matching point} (see
next section). In other words, this matching point is the outermost point in the
interior part and the innermost point in the envelope.

As the photosphere is not a part of the interior model, the first step is to
infer the current \teff by other means. We do this by setting up another
interpolation in the \staggrid, this time based on matching point temperature
\tma. By assuming the interior and envelope parts to have a common temperature
at the matching point, the \teff corresponding to this \tma can be evaluated
from the interpolation. With \teff established, we interpolate in the \staggrid
to obtain the gas pressure as a function of temperature -- in practice a scaled
pressure is used (discussed below). Then the gas pressure at the bottom of the
envelope is compared to the corresponding value predicted by the interior model
for its outer mesh point -- this is our boundary condition. Thus, in the
converged model the gas pressure of the outermost point in the interior matches
the value at the innermost point of the envelope -- of course the temperature
also matches by construction.

Outside the matching point we directly adopt $\pgas(T)$ from the interpolated
\tdes{} -- we refer to this part of the combined model as the \emph{appended
  envelope}. Here the density $\rho$ and the first adiabatic index \gam are
computed from the equation of state (EOS) used in the stellar evolution code.
The radius $r$ and mass $M(r)$ at each mesh point in the envelope are calculated
from hydrostatic equilibrium, and the photospheric radius of the star is
determined based on \teff and Stefan-Boltzmann's law.

In all of our models we use the OPAL opacities \citep{Iglesias1996} extended
with the low-temperature opacities from \citet{Ferguson2005}. To be consistent
with the simulations in the \staggrid we use the \citet[AGSS09,][]{Asplund2009}
solar composition. For the calculations, we employ the FreeEOS by A.~W.
Irwin\footnote{\url{http://freeeos.sourceforge.net/}} \citep{Cassisi2003}.

\subsection{The matching point}
\label{sec:method_match}

\begin{figure}
  \centering
  \includegraphics[width=\linewidth]{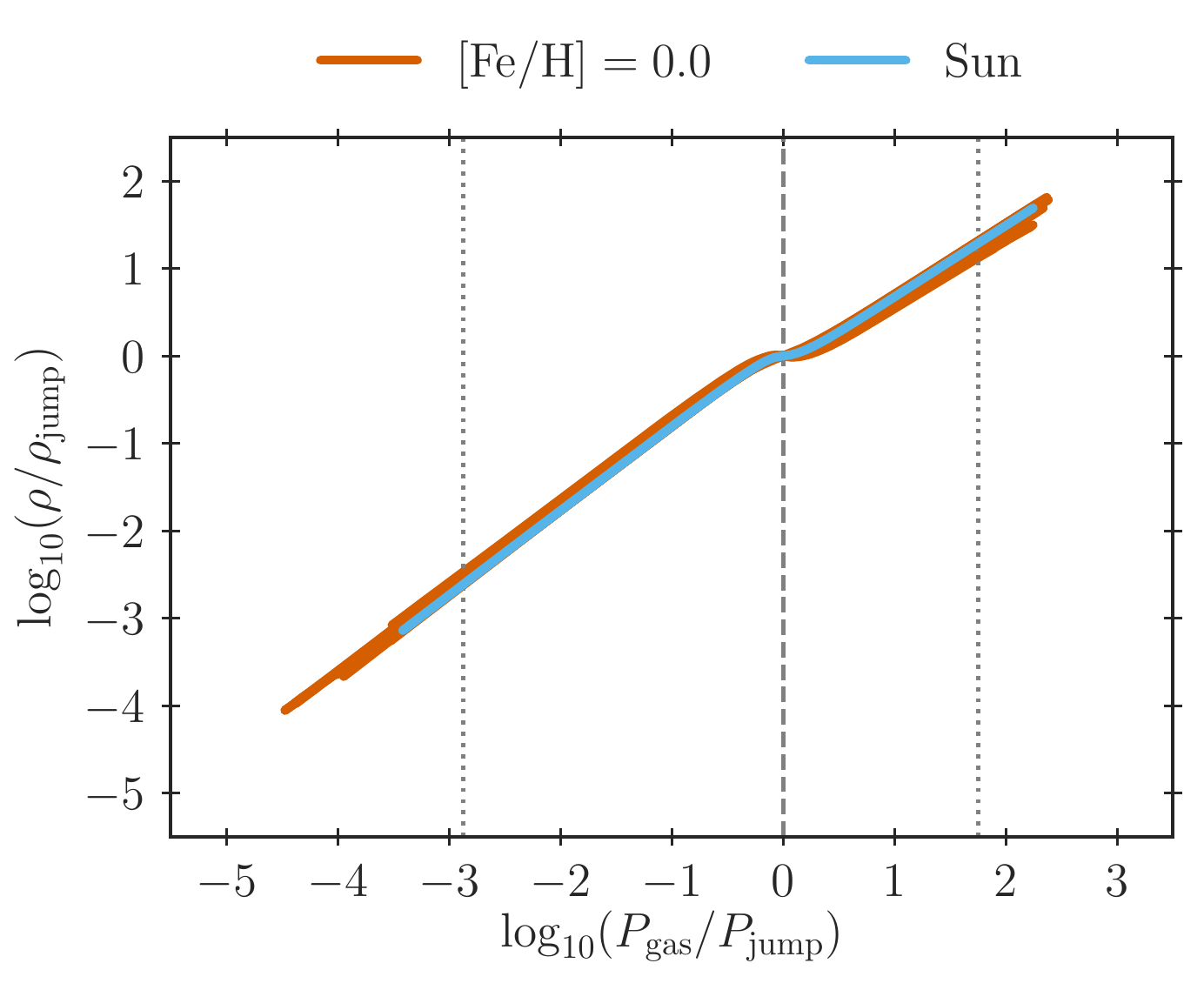}
  \caption{Scaled density stratification as a function of the scaled pressure
    for all 28 Stagger-grid simulations at solar metallicity. The stratification
    of the solar simulation is overlaid. The dotted lines mark the interpolation
    range. The dashed line highlights the density inflection point, the
    so-called density jump (details in text). For the equivalent plot, including
    all 199 relaxed simulations from the \staggrid, we refer to
    \citet[][Fig.~1]{Joergensen2019}.}
  \label{fig:densstrat}
\end{figure}

An important feature in the stratification of the 3D simulations is the minimum
in $\partial\log\rho / \partial\log \pgas$, which was called the \emph{density
  jump} in the nomenclature of \pp and \papI. Given the nature of this
near-surface feature, the term \emph{density inflection (point)} might be more
accurate; however, to avoid the confusion of modifying the nomenclature, we keep
the label \emph{jump} in the following. The pressure and density at this point
can be used to construct the \emph{scaled pressure} and \emph{scaled density},
which are the foundation of our method. The logarithm of these quantities are
shown for the simulations at solar metallicity from the \staggrid in
\fref{fig:densstrat}, where the density feature is also marked.

As introduced in \papI, the matching point between interior and appended \tde is
selected at a fixed scaled pressure. We introduce the quantity
\begin{equation}
  \label{eq:km}
  \km = \log_{10}\left(\frac{\pgas}{P_\mathrm{jump}}\right)_{\mathrm{matching \; point}} \; ,
\end{equation}
i.e, the value of the logarithm of the scaled pressure at the matching point is
dubbed \km. Our typical choice is $\km = 1.20$ -- which is near the right dotted
line in \fref{fig:densstrat} -- such that the pressure at the matching point is
$10^{1.20} \approx 15.8$ times higher than the pressure at the density feature
near the surface. The exact choice determines the depth and will influence the
produced model, which we will explore in \sref{sec:sun_depth}.

\section{Solar models}
\label{sec:sun}

In \papI we performed a solar calibration with our new implementation to obtain
a standard solar model (SSM) with \tdes appended \otf in the evolution. In the
following we will expand the discussion of the solar structure and evolution,
and address several aspects not treated in \papI.

\subsection{The equation of state}
\label{sec:sun_eos}

As mentioned above, temperature and pressure are taken directly from the
interpolated \tde in the appended part of the model, while we rely on the EOS
from \gar to supply the remaining quantities. \papI showed that the density is
recovered to very high accuracy. Another important quantity -- which we will
investigate in the following -- is the first adiabatic index $\gam = (\partial
\ln \pgas/\partial \ln \rho)_\mathrm{ad}$, which is not reproduced as accurately
and thus might affect the asteroseismic result.

We utilize the same solar calibration as described in \papI (Sec.~3), i.e., one
calibrated to yield $\teff = 5769\kel$ matching the \staggrid solar simulation.
The \tdes were used in the entire evolution. The matching point was at $\km =
1.20$ (see \eref{eq:km}) during the entire calculation, and the obtained
mixing-length parameter is $\mlt = 3.30$. Note that a direct comparison between
the mixing length used here and the standard mixing length used to characterize
the superadiabatic region in normal MLT is not meaningful. In the present case,
the role of the mixing-length parameter is to calibrate a tiny bit of
superadiabaticity below the matching region, and hence it is very sensitive to
the matching point and the details of the simulations. Thus, the actual
numerical value of \mlt{} are not important and in particular not relevant for
any other model.

The predicted stellar oscillation frequencies, $\nu_\mathrm{n\ell}$, are
calculated with the Aarhus adiabatic oscillation package
\citep[\adi,][]{Christensen-Dalsgaard2008a}. As discussed by \papI, \gam can
be used directly in the frequency computation because we (currently) neglect
turbulent pressure in our models constructed \otf
\citep{Rosenthal1999,Houdek2017}. In order to asses the impact of using the
adiabatic index from the EOS, we substitute the value computed by \gar, \gamm,
with the values taken directly from the interpolated \tde, \gammm, in our
calibrated solar model. We then recompute the oscillations, again using
\gam directly. As we are fully neglecting turbulent pressure, the different
\enquote{\gam cases} from \citet{Rosenthal1999} is not relevant.

\begin{figure}
  \centering
  \includegraphics[width=\linewidth]{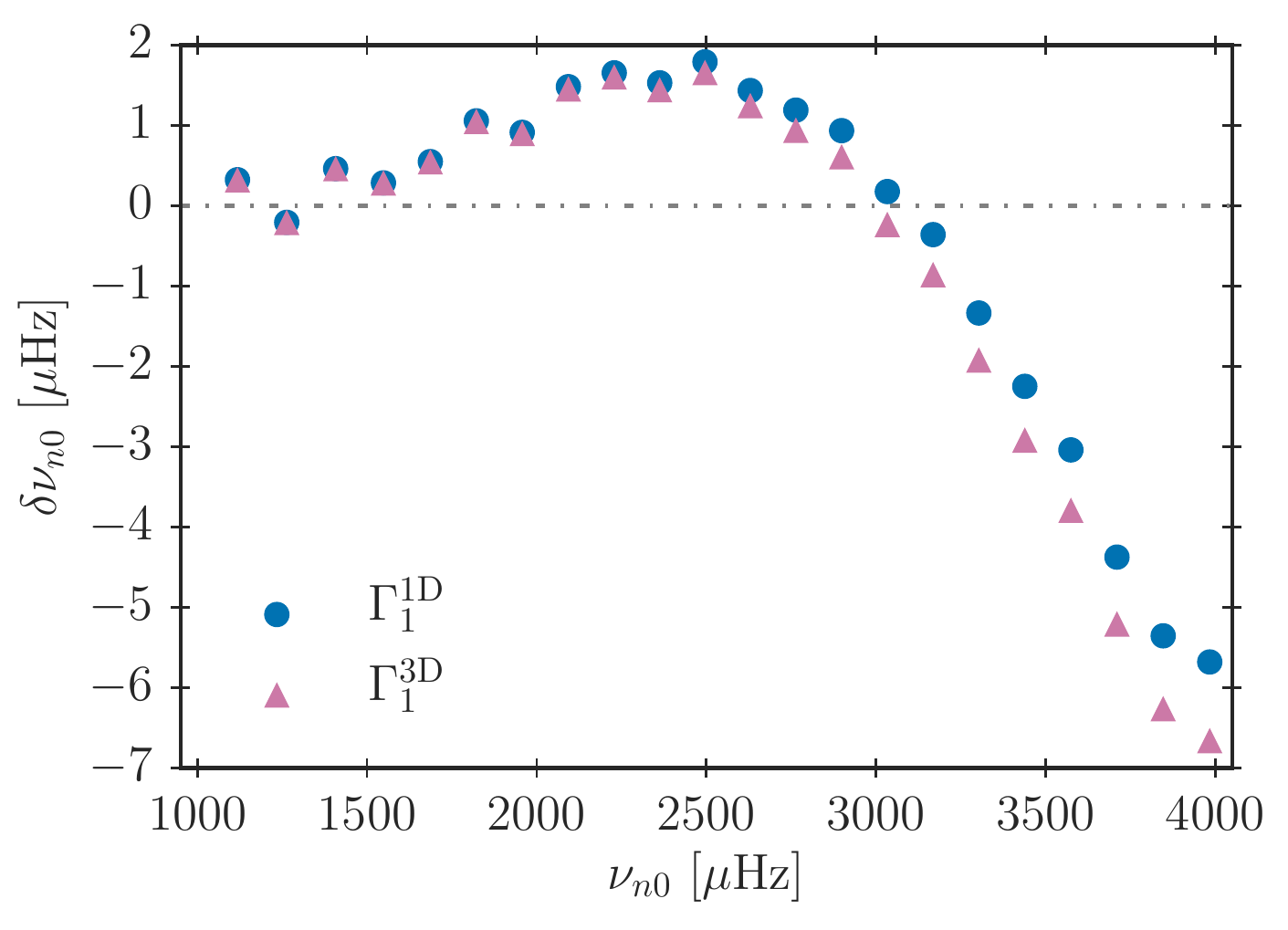}
  \caption{Frequency difference, $\delta \nu_\mathrm{n\ell}$, between
    radial modes ($\ell=0$) from model predictions and BiSON data. The
    predictions are from the the solar calibrated model from \papI using
    either \gam from \gar's EOS (\gamm) or directly from the interpolated \tde
    (\gammm).}
  \label{fig:gamma1}
\end{figure}

The comparison is shown in \fref{fig:gamma1}, as frequency differences compared
to observations from Birmingham Solar Oscillation Network
\citep[BiSON,][]{Broomhall2009,Davies2014}. To keep the comparison simple and to
not clutter the plot only radial modes ($\ell = 0$) are shown. As can be seen
from the figure, the effect is very small; the impact on the frequencies is
around $1 \muhz$ at the highest frequencies, and even less at the frequency of
maximum oscillation power, $\numax \simeq 3100\muhz$, and below.

\subsection{The matching depth}
\label{sec:sun_depth}

The choice of matching point (introduced in \sref{sec:method_match}) determines
the depth of the appended envelope and affects the parameters of the solar
calibration. In the following we perform ten different solar calibrations --
computed to match a standard solar $\teff = 5779\kel$ instead of the Stagger one
-- with different matching points in order to investigate to which extent the
matching point affects the obtained structure, evolution, and seismic results.
The different calibrations are denominated by the \km from \eref{eq:km}.

The selected matching point given by the scaled pressure is used in the entire
solar calibration routine, i.e., for the full evolution and not just in the
final solar model. This fixed scaled pressure of the matching point is by
construction constant throughout the evolution; however, for any given model in
the sequence, the radius coordinate \rma of the matching point -- valid only for
that particular model -- can be reported as well. For the final resulting solar
model this conversion can ease the discussion and especially in terms of
physical matching depth below the surface $\dma = R - \rma$.

\begin{figure}
  \centering
  \includegraphics[width=\linewidth]{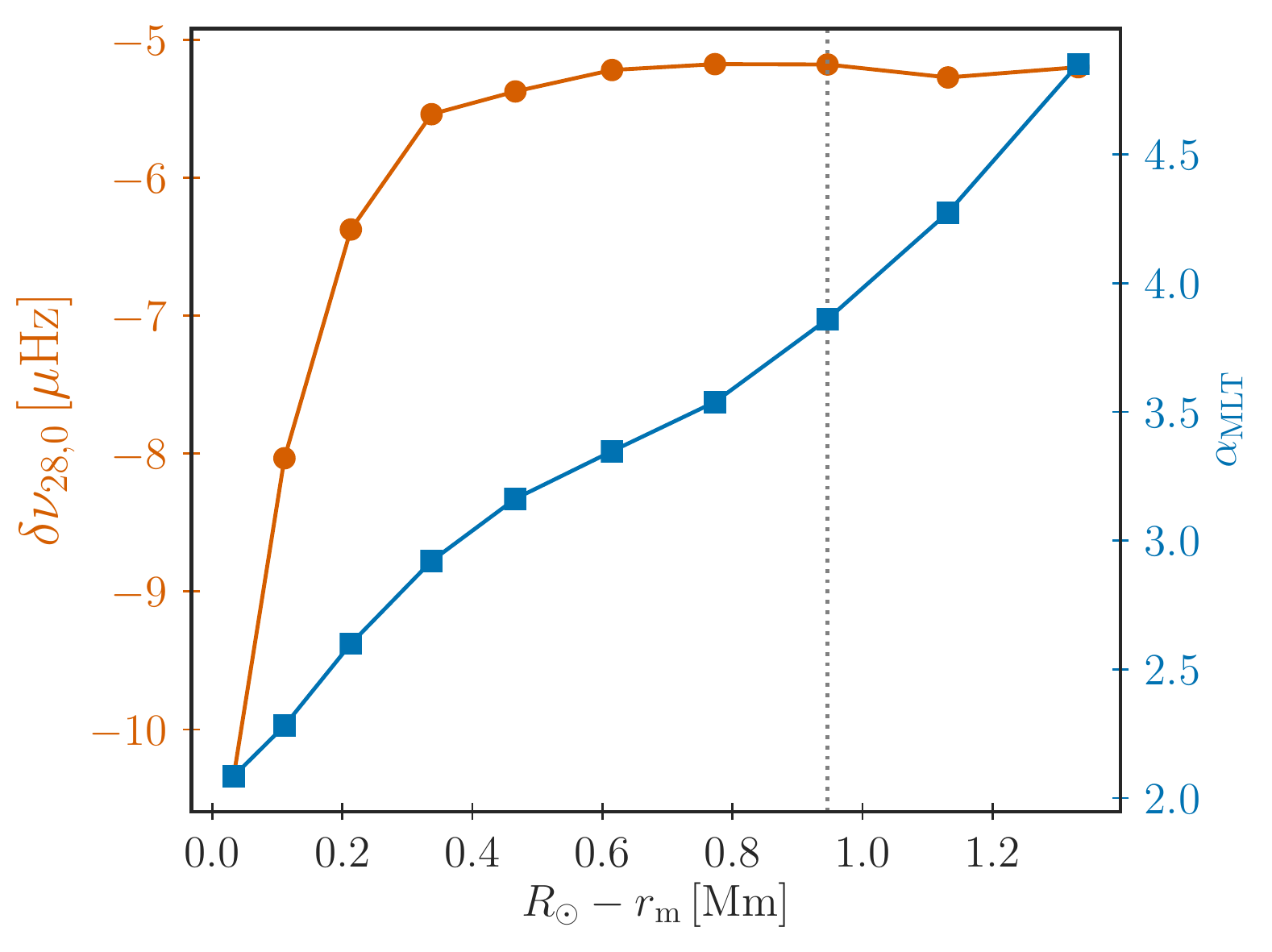}
  \caption{Results from the solar calibrations as a function of matching depth
    below the surface of the solar model. \emph{Left ordinate (orange circles):}
    Difference between BiSON data and model frequencies for the $\ell=0$, $n=28$
    mode. \emph{Right ordinate (blue squares):} The calibrated mixing length
    \mlt. The vertical dashed line marks the depth that correspond to $\km =
    1.20$, i.e., the scaled pressure at the matching point in
    \sref{sec:sun_eos}. The conversion to \km for several depths is presented in
    \fref{fig:evol_sun} and its caption.}
  \label{fig:alphafreq}
\end{figure}

Firstly, we analyse the stellar oscillations for each of the ten solar models.
Below a matching depth of $\dma \approx 0.6\mm$ -- which roughly corresponds to
the minimum in \gam near the surface -- we find the computed model frequency
differences to be virtually depth independent. Furthermore, when the matching
point is placed close to the surface, the obtained frequencies are very similar
to those obtained with a standard Eddington grey atmosphere. The difference
between the predicted model frequency and BiSON data for the $n=28$ radial
oscillation mode -- corresponding to roughly $4000\muhz$ -- is shown in orange
in \fref{fig:alphafreq} as a function of matching depth for all of the different
solar models.

Secondly, we investigate the mixing length parameter \mlt, which is an output of
the solar calibration. In order to correctly reproduce the solar surface
properties, higher values of \mlt are required when matching deeper below the
surface (i.e. at a higher scaled pressure). The corresponding values for each
matching depth are shown in blue in \fref{fig:alphafreq}, from which it is clear
that \mlt is found to monotonically increase with increasing \rma. A similar
result was obtained by \citet{Schlattl1997}, when appending mean structures of
2D envelope models. Also note that the calibration with the same matching depth
as used in the previous section and \papI ($\km = 1.20$) does not yield the same
mixing-length parameter, due to the different target \teff. The $10\kel$
difference changes the value from $\mlt = 3.30$ to $\mlt = 3.86$.

\begin{figure}
  \centering
  \includegraphics[width=\linewidth]{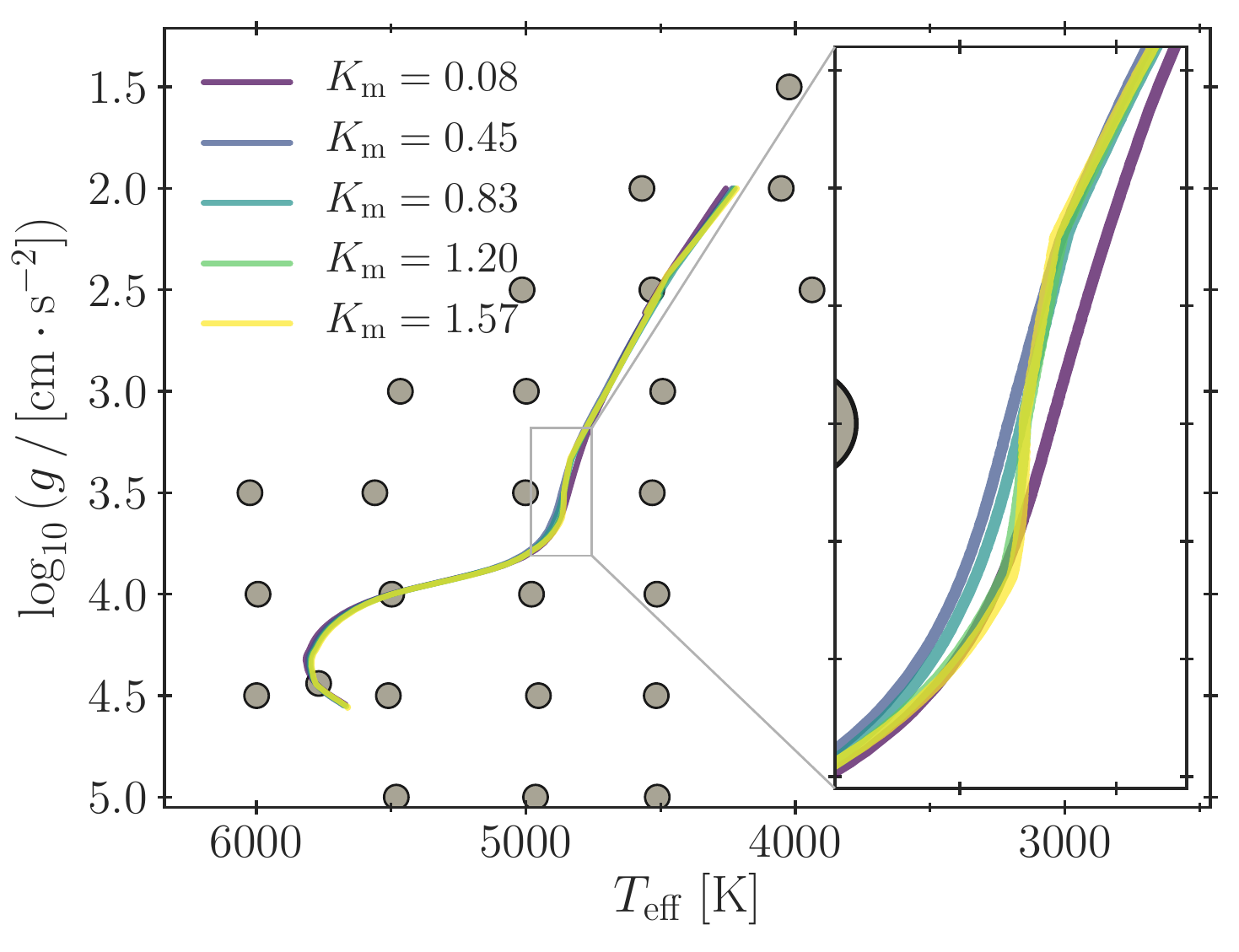}
  \caption{Evolution of the solar calibration models from the main sequence to
    the RGB. The ticks on the zoomed inset correspond to $100\kel$ and $0.1
    \dex$, respectively. The grey dots marks the \staggrid models at solar
    metallicity. The tracks are denoted by \km -- a larger value implies a
    larger matching depth. For the present Sun, the listed scaled pressures
    correspond to a depth of $0.11\mm$, $0.34\mm$, $0.61\mm$, $0.95\mm$ and
    $1.33\mm$, respectively (see \fref{fig:alphafreq}).}
  \label{fig:evol_sun}
\end{figure}

The influence of the matching point on the model's evolution is worth
investigating -- especially since the matching depth significantly alters the
calibrated value of \mlt. Therefore we calculated the evolution -- continuing up
the red-giant branch (RGB) -- of the solar calibrated models. The resulting
tracks for half of the cases (to not clutter the plot) are shown in
\fref{fig:evol_sun}. In the plot, the tracks are denoted by \km, which is kept
fixed for the entire evolution, and shown alongside the simulations from the
\staggrid used to obtain the interpolated \tde appended \otf.

From the figure it can be seen that the matching depth slightly alters the
position of the turn-off as well as the temperature evolution on the RGB, but
the effects are tiny. A more pronounced feature is the emerging kink at the
bottom of the RGB -- clearly visible from the zoom-in in \fref{fig:evol_sun} --
which we suspect to be a result of the interpolation. As discussed briefly
earlier and more extensively in \papI, we compute the effective temperature
\teff of the our models from interpolation in \logg and the temperature at the
matching point \tma (in \papI referred to as $\tma^{\mathrm{3D}}$). While the
\staggrid is almost regular in the $(\teff, \logg)$-plane, this is not the case
in the $(\tma, \logg)$-plane, due to the non-linear relationship between \teff
and \tma. This effect is shown in \aref{app:morph} for $\km = 1.20$. Moreover,
as the matching depth is changed, the simulation points move individually in
this parameter space, which causes the separation between them to change.

The result is that the larger the matching depth gets, the lower the resolution
in some regions of the $(\tma, \logg)$-plane is, which implies a higher risk for
interpolation errors in the determined \teff. As can be seen from
\fref{fig:evol_sun}, the evolutionary tracks show kinks on the RGB that become
more pronounced with increasing matching depth. Based on this, our method would
strongly benefit from a refinement of the \staggrid; specifically a few
additional 3D simulations with $\logg= 3.0-4.0$ and $\teff = 4500 - 5000\kel$.
We expand on the discussion of interpolation and grid resolution in
\sref{sec:evol_tracks}.

In order to fully take advantage of the 3D simulations, it is generally
desirable to place the matching point as deep within the nearly adiabatic region
as possible. As just mentioned, problems can however emerge in the post-main
sequence evolution if the matching is performed near the bottom of the
simulations. Thus, deciding on the matching depth is a compromise between these
two considerations. Until further 3D simulations are calculated, an intermediate
matching depth in the nearly adiabatic region is preferable -- such as the depth
used in \sref{sec:sun_eos} (identical to \papI) and in the following sections,
corresponding to $\km = 1.20$.

\subsection{The Trampedach grid}
\label{sec:sun_tramp}

To investigate the versatility of our method, we have repeated the analysis of
the solar model from \papI with a different set of 3D simulations: The grid from
\citet{Trampedach2013} consisting of 37 simulations at solar
metallicity.\footnote{We have excluded one of these simulations where some
  quantities were apparently missing from the averaged data.} In this section,
the interpolated \tdes are determined from the simulations in this grid.

For consistency, we calculate our models with the same non-canonical solar
mixture as this set of 3D simulations employs \citep[][Table 1]{Trampedach2013}
-- in which $Z/X = 0.0245$ -- as well as the specific atmospheric opacities from
\citet{Trampedach2014a,Trampedach2014b} provided by R.~Trampedach (priv. comm.).
The low-temperature opacities are merged with interior opacities from the
Opacity Project \citep[OP,][]{Badnell2005} for the same, identical composition.

The procedure for setting up the boundary conditions and appending the \tdes
\otf is identical to what is described in \sref{sec:method} and \papI. In the
nomenclature of earlier sections, the scaled pressure at the matching point
corresponds to $\km = 0.88$. For the present Sun, the pressure at the matching
point corresponds to a temperature of $\tma = 1.29 \times 10^{4}\kel$ and a
depth of $\dma = 0.64\mm$ below the photosphere. The appended envelope is hence
shallower than the envelope of the solar calibration model using the \staggrid
presented by \papI.

For comparison, we calculated a solar calibration with identical input physics,
but using a standard Eddington grey atmosphere -- labelled \emph{Edd.} in the
plots and table of this section. Moreover, the use of the grid from
\citet{Trampedach2013} and compatible input physics allows us to compare our
method to the work by \citet{Mosumgaard2018}, which is a different approach for
using information from 3D simulation in stellar evolution models. In this
procedure -- which also relies on interpolation in \teff and \logg -- the outer
boundary conditions are supplied by \ttaurel{}s extracted from the 3D
simulations by \citet{Trampedach2014a}, and the models include a variable
3D-calibrated \mlt from \citet{Trampedach2014b}. This specific solar model is
taken from \citet{Mosumgaard2018} and is denoted \emph{RT2014} in the following.

The resulting parameters from the three different calibrations are shown in
\tref{tab:suncal}. As found in \papI, our change in the outer boundary
conditions does not affect the surface helium mass fraction \ys nor the radius
of the base of the convection zone \rcz. For comparison, the results from
helioseismology are: $\ys = 0.2485\pm 0.0035$ \citep{Basu2004} and $\rcz = 0.713
\pm 0.001 \rsun$ \citep{Basu1997}. Regarding the mixing-length parameter, we
find the same as before and refer to the earlier discussion: that a direct
comparision of the two cases is not meaningful and that the actual values of
\mlt are not important.
\begin{table}
  \caption{Results from the different solar calibrations (here \tde is based on
    the \rtgrid; see text for further details). \mlt denotes the mixing length,
    \yi is the initial helium mass fraction, \zi is the initial fraction of
    heavy elements, \ys is the surface helium abundance, and
    $\frac{\rcz}{R_{\odot}}$ is the radius of the base of the convective
    envelope.}
  \label{tab:suncal}
  \begin{tabular}{lllllll}
    \hline
    Model & \mlt & \yi & \zi & \ys & $\frac{\rcz}{R_{\odot}}$ \\
    \hline
    Edd. & 1.71 & 0.2680 & 0.0201 & 0.2388 & 0.7122 \\[2pt]
    RT2014 & 1.82 & 0.2680 & 0.0201 & 0.2388 & 0.7122 \\[2pt]
    \tdes & 5.36 & 0.2678 & 0.0200 & 0.2387 & 0.7122 \\[2pt]
    \hline
  \end{tabular}
\end{table}

\begin{figure}
  \centering
  \includegraphics[width=\linewidth]{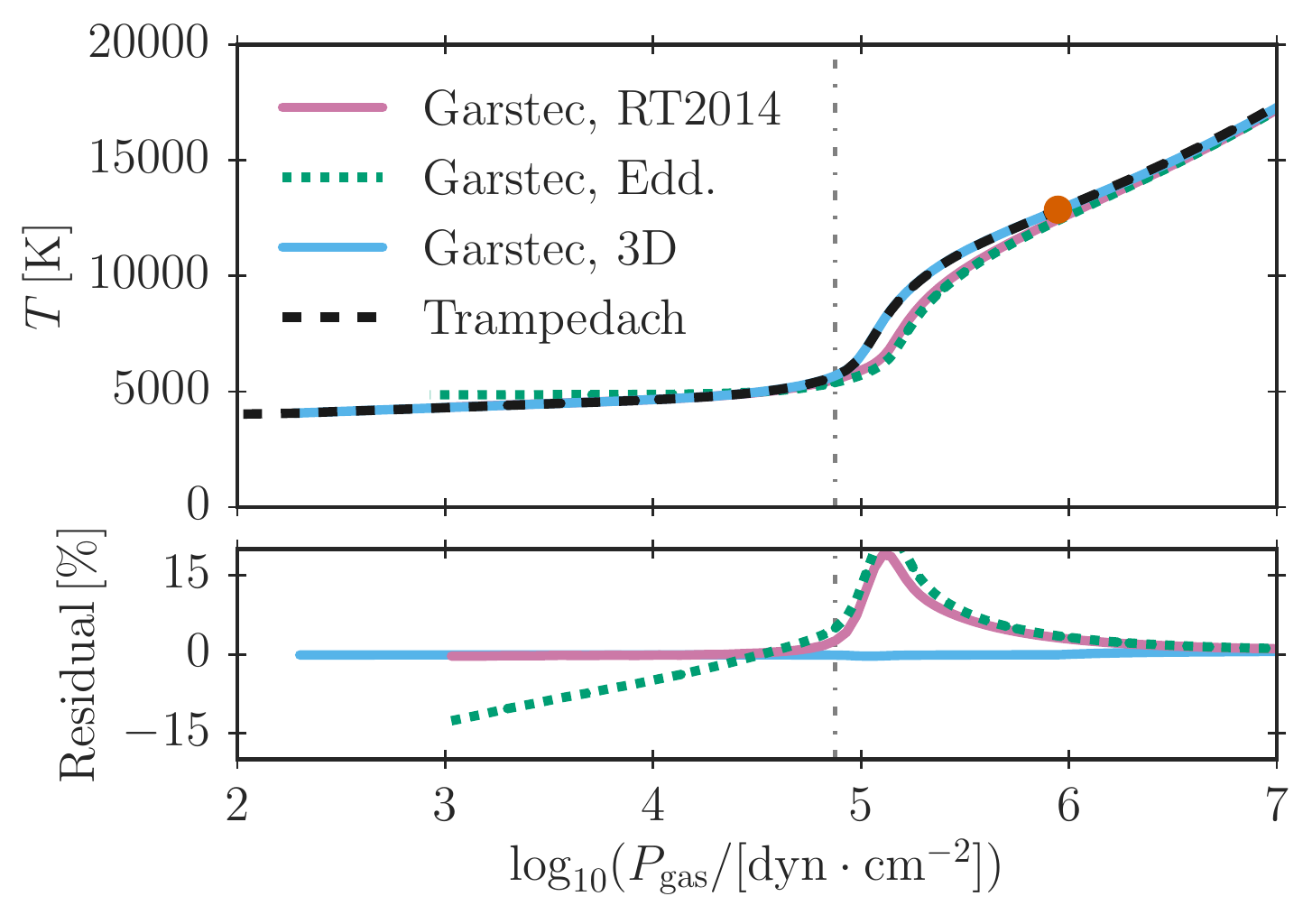}
  \caption{Comparison of the temperature as a function of gas pressure for the
    3D solar simulation in the Trampedach grid (\emph{Trampedach}) and the
    different solar calibrations from \gar: \emph{3D} is appending \tdes from
    the \rtgrid \otf, \emph{RT2014} is using a calibrated \mlt and \ttaurel from
    the same grid, and \emph{Edd.} is performed with a standard grey Eddington
    boundary and constant \mlt. The matching point of the \tde-model is marked
    with a red dot. The residuals are with respect to the 3D solar simulation,
    in the sense \emph{Trampedach} $-$ \emph{\gar}. The dashed-dotted grey line
    indicates the position of the photosphere. }
  \label{fig:Tvsp}
\end{figure}

A comparison of the temperature structure of the resulting solar models and the
3D solar simulation in the \rtgrid is shown in \fref{fig:Tvsp}. As can be seen
from the figure, our new method appending \tdes \otf reproduces the
stratification of the 3D simulation reliably throughout the envelope, which
agrees with the \staggrid results in \papI. It is clear that the Eddington grey
atmosphere is very different from the 3D hydro-simulation. The RT2014-solar
model using the 3D \ttaurel{}s and \mlt mimics the correct structure above the
photosphere, but deviates below -- a similar results was found by
\citet{Mosumgaard2018}.

\begin{figure}
  \centering
  \includegraphics[width=\linewidth]{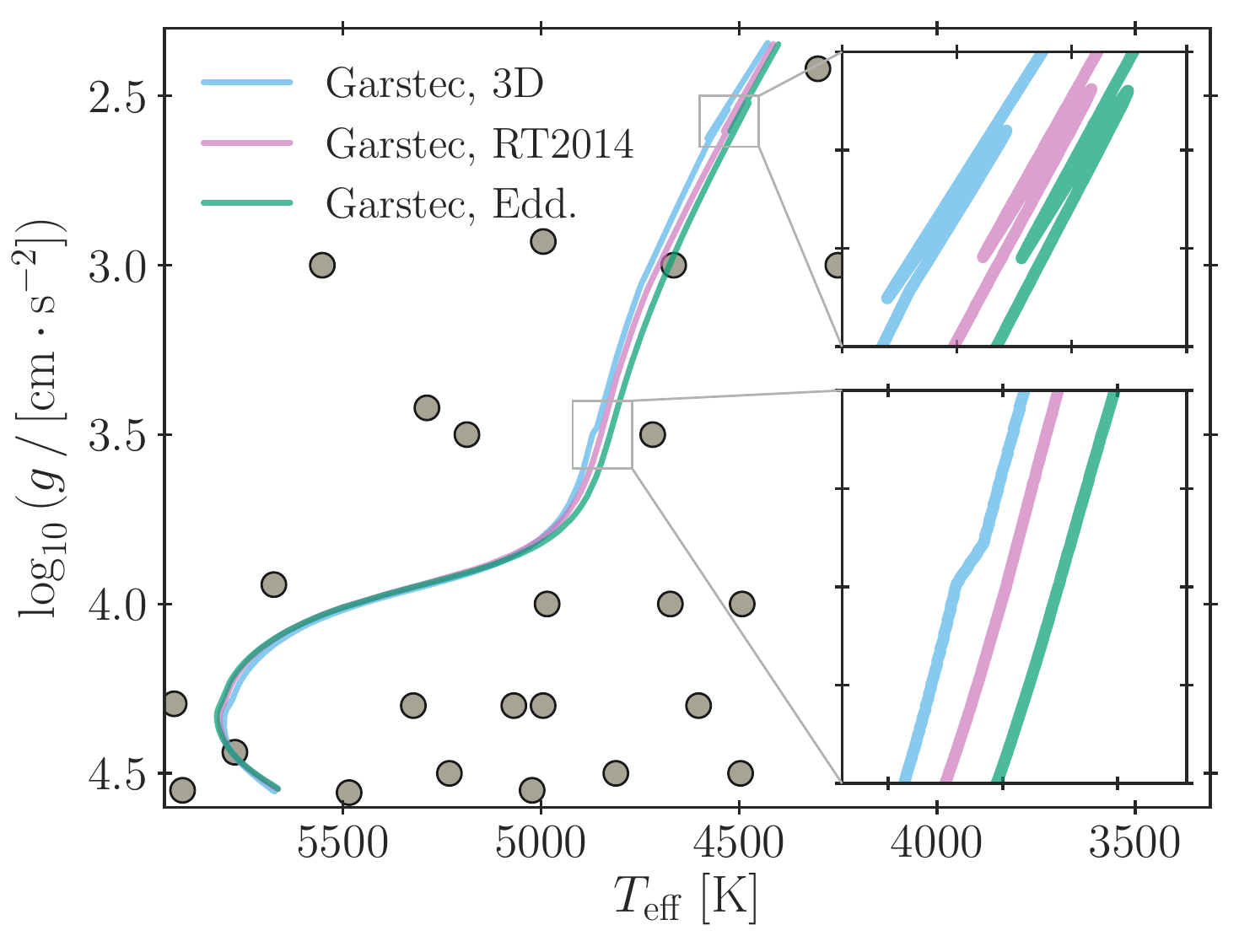}
  \caption{Evolutionary tracks for the solar models, with colors and labels
    corresponding to \fref{fig:Tvsp}. The ticks on the zoomed insets correspond
    to $50\kel$ and $0.05 \dex$, respectively. The grey dots show the location
    of a selection of the 3D simulations from the \rtgrid.}
  \label{fig:tracks1M}
\end{figure}

To asses the impact of using 3D information on the evolution, we continued the
tracks from the solar calibrations up the RGB. As shown by
\citet{Mosumgaard2018}, the use of the 3D calibrated \ttaurel and \mlt shifts
the RGB towards higher effective temperatures compared to the regular Eddington
case. As expected, we find the same qualitative trend for our implementation of
\tdes \otf, as can be seen in \fref{fig:tracks1M}, where the evolutionary tracks
are shown alongside the 3D simulations from the \rtgrid. In other words, models
with a standard Eddington grey atmosphere have systematically lower \teff on the
RGB. The difference between the track appending \tdes \otf and the Eddington
reference is around $\Delta\teff \simeq 55\kel$ at the RGB luminosity bump. This
difference is somewhere between the RGB temperature offset at solar metallicity
predicted by \citet{Tayar2017} and \citet{Salaris2018}; it is also around the
same magnitude as the shift found by the latter when changing between common
\ttaurel as boundary conditions for their models.

Moreover, \fref{fig:tracks1M} suggest that our new method leads to a kink at the
base of the RGB -- similar to what was seen in the previous section. We
attribute this again to interpolation errors due to the even lower resolution of
the \rtgrid in this region. The morphology of the kink is somewhat different --
more like a sharp break -- which likely stems from the irregular sampling of the
\rtgrid (compared to the almost uniform \staggrid). As pointed out in
\sref{sec:sun_depth}, such kinks call for a refinement of the currently employed
grids of 3D-envelopes -- regardless of the specific grid. In the following
section we will investigate the effects of the grid resolution in more detail.

As a final note, we repeated the full matching-depth analysis from
\sref{sec:sun_depth} for the \rtgrid. Bearing the slightly shallower Trampedach
simulations in mind, we observe the same qualitative behaviour for this grid as
we did for the \staggrid (shown in \fref{fig:alphafreq}). Specifically, \mlt
increases with increasing matching depth, reaching $\mlt = 17$ for a scaled
pressure corresponding to a matching depth of $\dma = 0.95\mm$ for the present
Sun. Regarding the frequencies as a function of matching depth, we observe the
same trend as before: Below a certain depth -- around $\dma = 0.5-0.6\mm$ which
is similar to what was seen for the \staggrid{} -- the frequencies are virtually
insensitive to the matching point. It should be noted that generally the
agreement between the models frequencies and observations are worse in this case
than for the \staggrid, as a result of the different opacities and chemical
mixture.

In the remainder of this paper, we will restrict ourselves to models that employ
the \staggrid rather than the \rtgrid, when appending \tdes \otf.

\section{Stellar Evolution}
\label{sec:evol}

To analyse the applicability of our procedure, we have produced a grid of
stellar models appending \staggrid \tdes \otf along the entire evolution. The
tracks are computed at solar metallicity with masses between $0.7\msun$ and
$1.3\msun$. In the calculations, the matching point is fixed to a scaled
pressure factor of $\km = 1.20$, which is the same as in \sref{sec:sun_eos} and
\papI. The tracks use a fixed $\mlt = 3.86$ from the solar calibration with the
corresponding depth from \sref{sec:sun_depth}. A list of the input physics is
provided in the final paragraph of \sref{sec:method}.

\subsection{Evolutionary tracks and grid resolution}
\label{sec:evol_tracks}
\begin{figure}
  \centering
  \includegraphics[width=\linewidth]{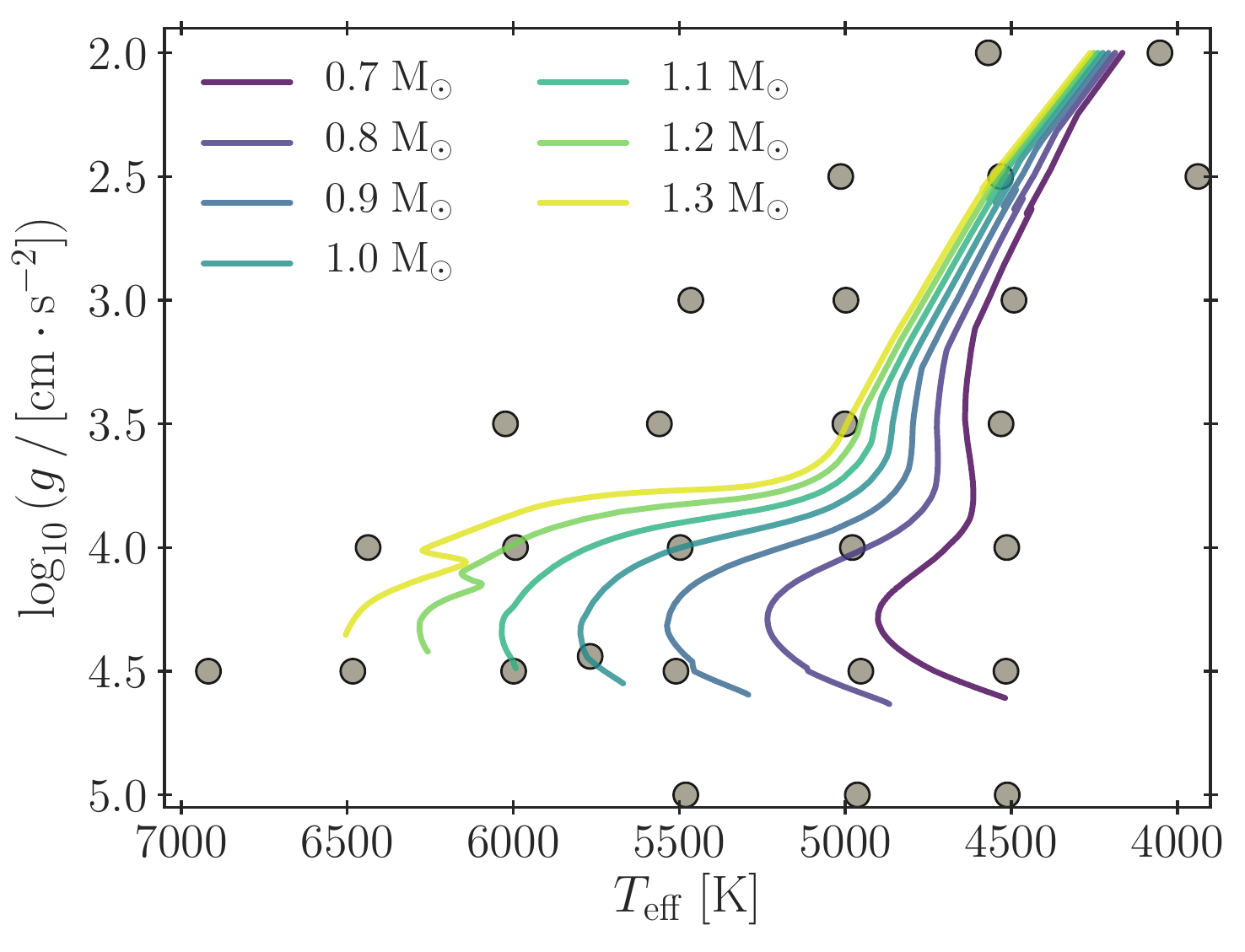}
  \caption{Evolutionary tracks for models with different stellar masses. The
    grey dots show the location of the \staggrid models with solar metallicity.}
  \label{fig:tracks220dif}
\end{figure}

A selected subset of the evolutionary tracks spanning the entire mass range of
our grid is shown in \fref{fig:tracks220dif} up to $\logg = 2.0$. As can be seen
from the figure, the evolutionary sequences are generally well behaved, but show
different kinks (or changes in slope) -- especially visible on the RGB, but not
at the same location for the different tracks.

The most prominent of these features are located between $\logg=3.5$ and
$\logg=3.0$, which is in the same region of the Kiel diagram where difficulties
emerged for the solar model tracks in \fref{fig:evol_sun}. Similar kinks can be
seen for the majority of the tracks between $\logg=2.5$ and $\logg=2.0$, and
also in the main sequence for the $0.8\msun$ and $0.9\msun$ evolution. All of
the cases are correlated with larger gaps in the \staggrid -- and also with
movement of the simulation footpoints in the (\tma, \logg)-plane (see
\aref{app:morph}). Thus, the bends generally occur on the virtual line between
two of the simulations, i.e., when the tracks move to a different zone in the
triangulation-based interpolation scheme (in either one of the two parameter
spaces). Specifically, it seems to be a problem with the sampling of the
underlying grid of 3D simulations.

To investigate the influence of the grid sampling, we performed numerous tests
of the triangulation and interpolation. We modified the grid used by our
routines in \gar; specifically we tried degrading the grid by strategically
removing some of the Stagger models. We also employed the code from \pp to
compute new interpolated envelopes in the gaps (e.g. at $\teff = 4775$, $\logg =
3.75$) to artificially refine the grid. All of the tests confirm the actual grid
sampling to clearly affect the morphology of the RGB-kink: making the break
smoother/sharper and more/less pronounced. The effect of the sampling in the
different interpolation planes are discussed in \aref{app:morph}. To sum up, we
need a denser grid of 3D hydro-simulations in order to produce smoother
evolutionary sequences.

\fref{fig:tracks220dif} illustrates the well-known effect that tracks are much
closer to each other in temperature for the later evolutionary stages with lower
surface gravities; at the zero-age main sequence (ZAMS) they span more than
$2000\kel$, but this gets narrower moving up the RGB and the extent is less than
$100\kel$ at $\logg = 2.0$. In other words, during the main sequence the
evolutionary tracks of different initial mass is spread across the entire grid,
while the effective resolution is significantly reduced for red giants, with
only a few simulations to cover the entire mass range. Another interesting
observation is that the separation between the tracks decreases going up the
RGB, whereas they are mostly parallel for standard evolution with an Eddington
atmosphere. A potential future line of investigation would be to determine
whether this is a true effect, or if it is due to a deficit in the low \logg
simulations, or a result of the RGB grid resolution.

Looking at the figure, a final important thing to keep in mind is that the
applicability of our method is strongly determined by the parameter space
covered by the 3D hydro-grid -- both in terms of mass range and how far up the
RGB the tracks can extend.

\subsection{Structure at different evolutionary stages}
\label{sec:evol_stages}

Some of the evolutionary tracks contain models, where \teff and \logg correspond
to one of the existing \staggrid simulations -- this is directly visible from
\fref{fig:tracks220dif}, where some of the selected tracks pass through a dot.
This facilitates an easy comparison between the obtained structure from the
appended \tde in the stellar evolution model and the original 3D simulation. In
other words, we want to verify that the direct output from our stellar structure
model -- including the quantities derived from the EOS -- is consistent with the
underlying full 3D simulations.

We have performed several of such comparisons, which are listed in
\tref{tab:match} alongside the deviation in \logg (at the matching point) and
\teff between the model and corresponding 3D simulation. The matching is within
roughly $0.8\kel$ and $10^{-3} \, \mathrm{dex}$, resulting in relative
deviations at the $10^{-4}$ level or better. Especially the high precision in
surface gravity at the matching point is important, as the interpolation is very
sensitive to \logg. In the table, we have adopted the nomenclature from the
\staggrid to label the models: As an example, the model named \texttt{t50g35}
has $\teff = 5000\kel$ and $\logg = 3.5$.
\begin{table}
  \centering
  \caption{Difference and relative deviation in \teff at the
    photosphere and \logg at the matching point, between the 1D stellar model
    and corresponding 3D simulation (in the sense 1D $-$ 3D). The nomenclature
    specifies the surface parameters of the simulations: E.g. the model
    denoted \texttt{t55g45} has $\teff = 5500\kel$ and $\logg = 4.5$.}
  \label{tab:match}
  \begin{tabular}{lrrrr}
    \hline
    Sim.             &  $\delta \, \teff$ [K]  &  $\delta \, \teff / \teff$ &  $\delta \logg$ [cgs]  &  $\delta \logg / \logg$  \\
    \hline
    \texttt{t45g25}  &  $-0.535$               &  $-1.19 \times 10^{-4}$    &  $-5.10 \times 10^{-5}$ &  $-2.04 \times 10^{-5}$  \\
    \texttt{t50g35}  &  $-0.248$               &  $-4.96 \times 10^{-5}$    &  $7.79 \times 10^{-4}$  &  $2.23 \times 10^{-4}$   \\
    \texttt{t55g40}  &  $0.600$                &  $1.09 \times 10^{-4}$     &  $7.83 \times 10^{-4}$  &  $1.96 \times 10^{-4}$   \\
    \texttt{t55g45}  &  $0.810$                &  $1.47 \times 10^{-4}$     &  $1.56 \times 10^{-3}$  &  $3.47 \times 10^{-4}$   \\
    \texttt{t60g40}  &  $-0.360$               &  $-6.00 \times 10^{-5}$    &  $-9.93 \times 10^{-4}$ &  $-2.48 \times 10^{-4}$  \\
    \hline
  \end{tabular}
\end{table}

\begin{figure}
  \centering
  \includegraphics[width=\linewidth]{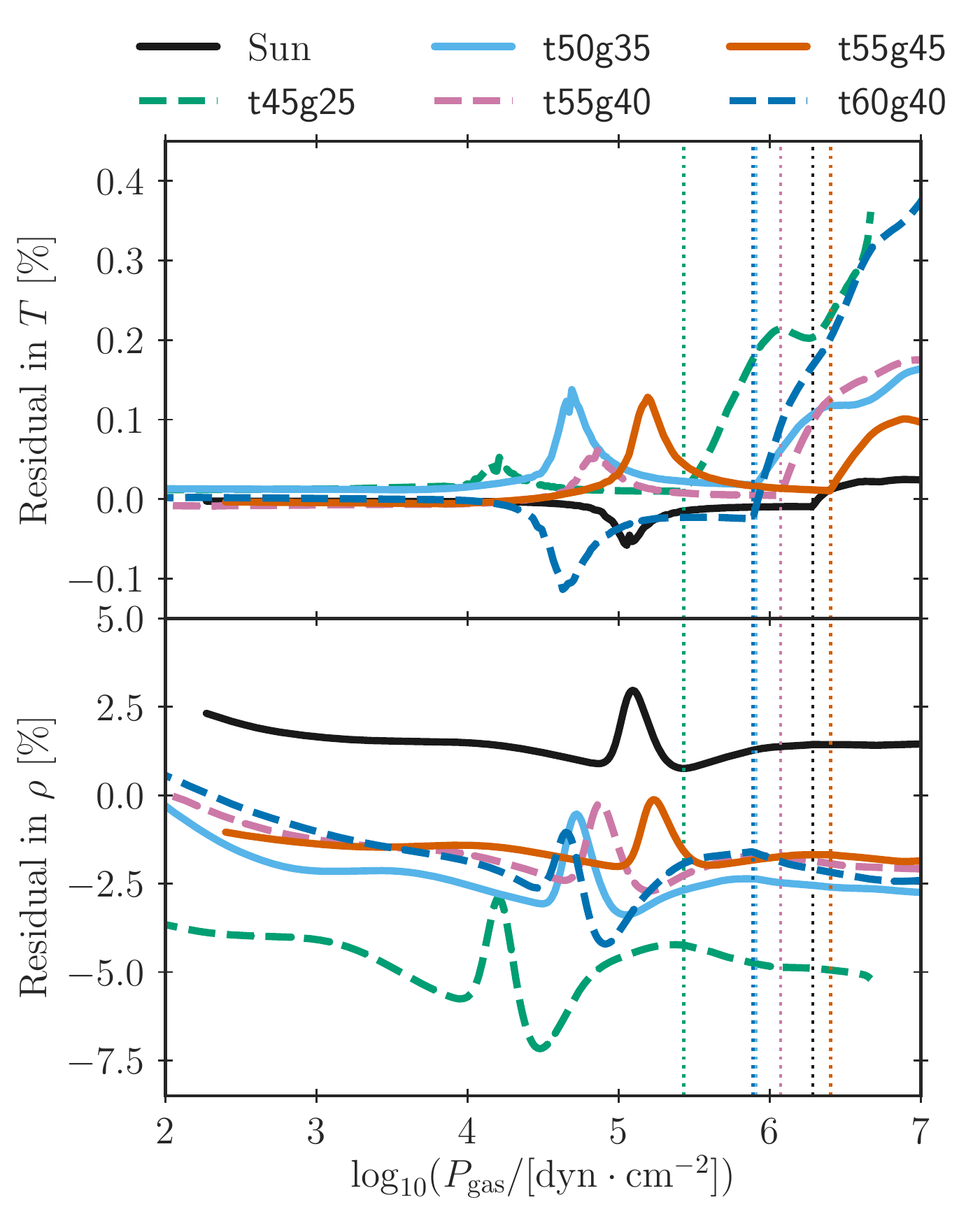}
  \caption{\emph{Upper panel:} Comparison between the temperature stratification
    determined by \gar and the corresponding \staggrid simulation at solar
    metallicity for the selected cases of \tref{tab:match}. The nomenclature is
    the same. The residuals are calculated as \emph{3D simulation} $-$
    \emph{\gar}. For each set of surface parameters, a vertical dotted line with
    the corresponding color indicates the location of the matching point.
    \emph{Lower panel:} As the upper panel but comparing the density
    stratification.}
  \label{fig:resrhoT}
\end{figure}

The resulting residuals in temperature and density as a function of gas pressure
for five cases are shown in \fref{fig:resrhoT} (using the same nomenclature),
where the comparison of the Sun is added for reference.
As can be seen from the top panel of the figure, our method reproduces the
temperature stratification with high accuracy throughout the $(\teff,
\logg)$-plane, with residuals below $0.2\%$ above the matching point. The
smallest residuals are seen for \texttt{t45g25}, which has the best match to the
simulation surface parameters. But generally we find no clear trends in the
residuals with the deviation in matching parameters from \tref{tab:match}.

Regarding the density -- shown in the bottom panel of \fref{fig:resrhoT} -- the
procedure works particularly well for the main-sequence and subgiant stars in
our sample, i.e., excluding the giant \texttt{t45g25} (discussed separately
below). The residuals are within a few percent, which is very similar to the
levels seen for the solar model in \papI. As for the temperature, we see no
correlation of the residuals with how well the models match. Because the
interpolation is entering only through the temperature stratification, the
density residuals are not directly reflecting (additional) interpolation errors.
As was also argued in \papI, the discrepancies partly reflect the difference in
EOS, but also the non-linearity of the thermodynamic quantities. Specifically,
we derive the density from the mean of \pgas and $T$, i.e., $\rho (\mean{\pgas},
\mean{T})$. Regardless of the EOS, this density is not expected to correspond
exactly to the actual (geometrical) mean of the density, $\mean{\rho(\pgas,
  T)}$, in a 3D hydro-simulation, because the density is a non-linear function
of pressure and temperature \citep{Trampedach2014a}.

For the more evolved giant shown (\texttt{t45g25}), the residuals in density
between the \staggrid simulation and outer parts of the stellar evolution model
are somewhat larger -- even though the level of the temperature residuals are
not larger. The cause must be the above-mentioned difference in EOS and
thermodynamic non-linearity, which clearly play a much larger role in the
red-giant regime. Another possible contribution is a composition effect;
however, this is unlikely to constitute the full part, because compared to the
initial model in the sequence the difference is less than $0.01$ in \feh and
around $0.02$ in surface helium.

To sum up, the overall implementation is performing less well in the red-giant
region of the parameter space. This is in line with the findings by \pp, namely
that the residuals in the patching grows for the hottest and the most evolved
stars. Thus, to fully utilize the potential of out method, more 3D simulations
are required.

\section{Asteroseismic Application}
\label{sec:kepler}

We want to investigate how our new procedure alters the results obtained from an
asteroseismic analysis compared to a reference case. The selected stars must
have an \teff and \logg inside the \staggrid, and as we currently do not
interpolate in metallicity, we are restricted to stars with a composition
consistent with solar. Moreover, the method is expected to primarily perform
well for cold main-sequence stars, as discussed above. Based on these
restrictions, we have selected two stars from the \kep asteroseismic \leg sample
\citep{Lund2017,Aguirre2017} with large frequency separations around $\dnu \sim
155\muhz$: KIC~9955598 ($\dnu = 153.3\muhz$) and KIC~11772920 ($\dnu =
157.7\muhz$). The stellar parameters resulting from the fit (described in the
next section) are listed in \tref{tab:fit}.

For the grid-based modelling analysis, two sets of stellar models have been
computed: One appending \tdes from the \staggrid \otf and a reference grid with
an Eddington grey atmosphere. The model grids contain the same microphysics as
listed in \sref{sec:method}, but do not include microscopic diffusion. The grid
appending \tde \otf uses $\mlt = 3.86$ (the $\km = 1.20$ solar calibration from
\sref{sec:sun_depth}), while the Eddington solar calibration yields $\mlt =
1.80$.

The grids have been calculated with \gar and span the mass range $M = 0.8 -
0.95\msun$ in steps of $0.001\msun$. For all models in the grids, \adi
\citep{Christensen-Dalsgaard2008a} has been utilised to calculate individual
oscillation frequencies.

\subsection{Determined stellar parameters}
\label{sec:kepler_params}

To compare the observations to the calculated stellar models we have used \bas
\citep[\textbf{BA}yesian \textbf{ST}ellar
\textbf{A}lgorithm,][]{Aguirre2015,Aguirre2017}, which utilises both classical
observables and asteroseismic data. Based on the observed quantities, the
likelihood of all models in the grid is determined, and probability
distributions and correlations constructed for the desired parameters. The
reported values are the medians from these distributions with the $68.3$
percentiles as corresponding uncertainties.

One way of using the asteroseismic data is to compare the observed individual
oscillation frequencies $\nu_{n,l}$ to those computed from the models. Usually
this approach requires an analytical prescription to correct for the surface
effect \citep[e.g.,][]{Kjeldsen2008,Ball2014}. It should be noted that the exact
shape of this correction is not known for our new stellar models appending \tdes
\otf, where the surface effect has been partly eliminated.

Another option is to use combinations of frequencies instead of the individual
frequencies; specifically the frequency separation ratios defined as
\citep{Roxburgh2003}:
\begin{align*}
  r_{01}(n) &= \frac{\nu_{n-1,0}-4\nu_{n-1,1} + 6 \nu_{n,0} - 4\nu_{n,1} + \nu_{n+1,0}}{8(\nu_{n,1}-\nu_{n-1,1})}, \\
  r_{10}(n) &= \frac{-\nu_{n-1,1} + 4\nu_{n,0} - 6\nu_{n,1} + 4\nu_{n+1,0} - \nu_{n+1,1}}{8(\nu_{n+1,0}-\nu_{n,0})}, \\
  r_{02}(n) &= \frac{\nu_{n,0}-\nu_{n-1,2}}{\nu_{n,1}-\nu_{n-1,1}} \; .
\end{align*}
These frequency separation ratios have been shown to be less sensitive to the
outer layers and primarily probe the interior of the star \citep[see,
e.g.,][]{Oti2005,Roxburgh2005}, and thus eliminate the need for a correction of
the surface term. Typically the ratios are combined into a set of observables,
e.g.:
\begin{equation*}
  r_{010} = \left\{ r_{01}(n),\; r_{10}(n),\; r_{01}(n+1),\; r_{10}(n+1),\; \dots \right\} \; .
\end{equation*}

To test the consistency of our new models, we used BASTA to estimate the stellar
properties based on a fit to the spectroscopic temperature and the frequency
ratios $r_{010}$ and $r_{02}$.\footnote{The use of the $r_{010}$ in combination
  with $r_{02}$ was disputed by \citet{roxburgh18a}, due to the risk of
  overfitting the data, and instead suggested using the single series $r_{102}$
  (or $r_{012}$). Using the $r_{012}$ in BASTA, we obtain parameters fully
  consistent with the \enquote*{standard} ratios fit.} In the current context,
the agreement between the two fits and not the actual parameter values is our
primary concern. However, to guide the discussion the inferred stellar
parameters from both sets of models are listed in \tref{tab:fit}.
\begin{table*}
	\centering
	\caption{Parameters of the modelled \kep stars, appending \tde \otf
    (\emph{Stag.}), or using Eddington grey atmospheres (\emph{Edd.}) as
    boundary conditions. inferred using \bas (details in the text). The listed
    values correspond to the median of the obtained probability distributions
    from \bas and the uncertainties denote $68.3\,\%$ bayesian credibility
    intervals.}
	\label{tab:fit}
	\begin{tabular}{lccccccccccccccccccc}
		\hline
    KIC      &  Model  &  \teff [K]          &  \logg [cgs]                &  Mass [\msun]             &  $R_{\mathrm{phot}}$ [\rsun]   &  Age [Myr]           \\
    \hline
    9955598  &  \emph{Edd.}   &  $5572^{+13}_{-13}$  &  $4.4983^{+0.0011}_{-0.0012}$  &  $0.897^{+0.005}_{-0.005}$  &  $0.8839^{+0.0021}_{-0.0019}$  &  $6997^{+360}_{-349}$ \\[6pt]  
    9955598  &  \emph{Stag.}  &  $5584^{+10}_{-10}$  &  $4.4989^{+0.0011}_{-0.0012}$  &  $0.899^{+0.004}_{-0.005}$  &  $0.8840^{+0.0019}_{-0.0019}$  &  $6944^{+352}_{-322}$ \\[6pt]  
    11772920 &  \emph{Edd.}   &  $5423^{+15}_{-15}$  &  $4.5061^{+0.0013}_{-0.0013}$  &  $0.849^{+0.005}_{-0.006}$  &  $0.8520^{+0.0023}_{-0.0023}$  &  $9874^{+528}_{-499}$ \\[6pt]  
    11772920 &  \emph{Stag.}  &  $5449^{+16}_{-16}$  &  $4.5069^{+0.0013}_{-0.0013}$  &  $0.852^{+0.006}_{-0.006}$  &  $0.8529^{+0.0023}_{-0.0022}$  &  $9752^{+510}_{-506}$ \\[6pt]  
		\hline
	\end{tabular}
\end{table*}

In general, the resulting parameters from the grid of \tde models and the grid
of Eddington models show good agreement. The effective temperature is
particularly interesting, as we know from an earlier section that the \teff
evolution can be different between the two sets of models. However, not much is
predicted to change on main sequence; as expected, for KIC~9955598 the two
values are within the uncertainties of each other, while for KIC~11772920 the
quoted uncertainty bands in \teff overlap. For both stars, for all of the
remaining parameters -- mass, radius and age -- the agreement between the two
grids is even better, and within half a standard deviation of each other.

After verifying the consistency with the non-surface dependent separation
ratios, we repeated the procedure using instead the individual oscillation
frequencies. We assume the two-term surface correction from \citet{Ball2014} and
fit the stars using the same two sets of models. The inferred parameters from
this analysis are not shown; however they are similar to the presented results
from the fit to the $r_{010}$ and $r_{02}$ ratios. The determined parameters
from the Eddington and the \tde grid show the same level of agreement as above,
i.e., less than half a standard deviation for all of the parameters except
\teff. Moreover, all of the fits to the same star using the different sets of
asteroseismic observables are internally consistent, too.

\subsection{The surface effect}
\label{sec:kepler_surface}

Besides estimating the stellar parameters, we want to investigate the impact on
the important asteroseismic surface effect. In order to isolate the surface
term, we add an additional assumption to our fitting: The model must match the
observed lowest order $\ell = 0 $ mode within $3\sigma$. From \bas we can get
the full stellar model corresponding to the point with the highest assigned
likelihood -- given our extra assumption -- also known as the best-fitting model
(BFM). By comparing the BFM from each grid, we can investigate if our new models
alters the individual oscillation frequencies.

\begin{figure}
  \centering
  \includegraphics[width=\linewidth]{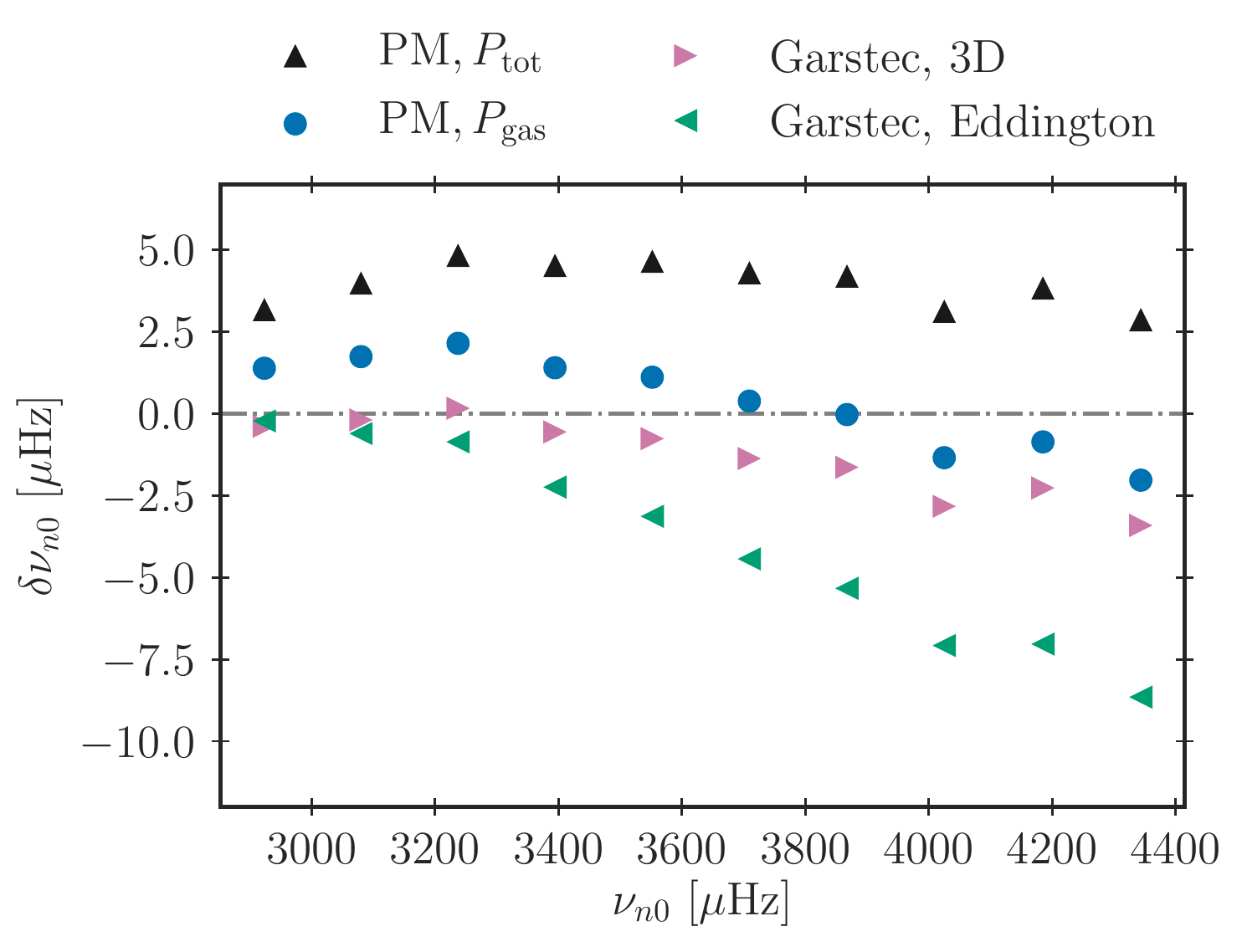}
  \caption{Frequency differences between predicted $\ell = 0$ model frequencies
    and observations for KIC~11772920. The plot includes the best-fitting model
    with an Eddington grey atmosphere and with \staggrid \tdes \otf. Based on
    the latter, we have furthermore constructed two patched models (PM): one,
    for which the total pressure ($P_\mathrm{tot}$) enters the computation of
    the depth scale, using hydrostatic equilibrium, and one, for which only the
    gas pressure \pgas is used to compute $r$.}
  \label{fig:KIC11772920_indfreq}
\end{figure}

In \fref{fig:KIC11772920_indfreq}, the BFM-comparison for KIC~11772920 is shown
in the form of frequency difference with respect to the observations. Looking at
the figure, it is clear that the frequencies of our new model appending \tdes
\otf deviates less from the observations, without the need of a surface
correction.

The figure likewise contains two patched models (PM) constructed following the
procedure described by \papI and \pp. The base model for the present patching
exercise is the \gar model employing our new implementation. In each case we
have substituted the outer layers of this model with the full \td-structure of
an interpolated \staggrid envelope with the same \teff and \logg. The first case
-- denoted as \enquote{PM, $P_{\mathrm{tot}}$} in the figure -- is taken as it
is, i.e., it includes turbulent pressure in the patched layers. In the second
case, which is dubbed \enquote{PM, $P_{\mathrm{tot}}$}, the depth scale of the
patched \td-envelope is recalculated solely based on the gas pressure.

Since our \gar implementation neglects turbulent pressure and infers $\rho$ and
\gam from the EOS, especially the first case substitution is expected to alter
the structure and affect the model frequencies (cf. e.g. \papI and
\citealt{JoergensenWeiss2019}). To facilitate a meaningful comparison with the
frequencies from stellar evolution models, we do not include the contribution
from turbulent pressure in the oscillation equations for this PM, i.e., we
assume \enquote{gas \gam approximation} in the nomenclature of
\citet{Rosenthal1999}. For an elaboration of this see \papI (Sec. 3.2) and
\citealt{Houdek2017}.

From \fref{fig:KIC11772920_indfreq} we see that including the turbulent pressure
in the patched exterior give rise to model frequencies which are $4-7\muhz$
lower than the frequencies of the underlying \gar model. Note that the so-called
\emph{modal effects} -- including non-adiabatic contributions in the computation
of the mode frequencies in the separate pulsation code (\adi) -- are not
included in the current treatment, but would still play a significant role for
the remaining discrepancy \citep{Houdek2017}.

When recomputing the depth scale of the patched \td-envelope purely based on the
gas pressure, this mismatch in the oscillations is reduced to $\lesssim 2\muhz$.
This illustrates the importance of taking turbulent pressure properly into
account. The remaining discrepancy between the PM and the model that has been
obtained, using our new implementation, may partly be attributed to a mismatch
in the stratification of $\rho$ or \gam --- that is, frequency differences may
be attributed to the EOS or assumptions made by the EOS. Furthermore, this
discrepancy may partly reflect interpolation errors.

\begin{figure}
  \centering
  \includegraphics[width=\linewidth]{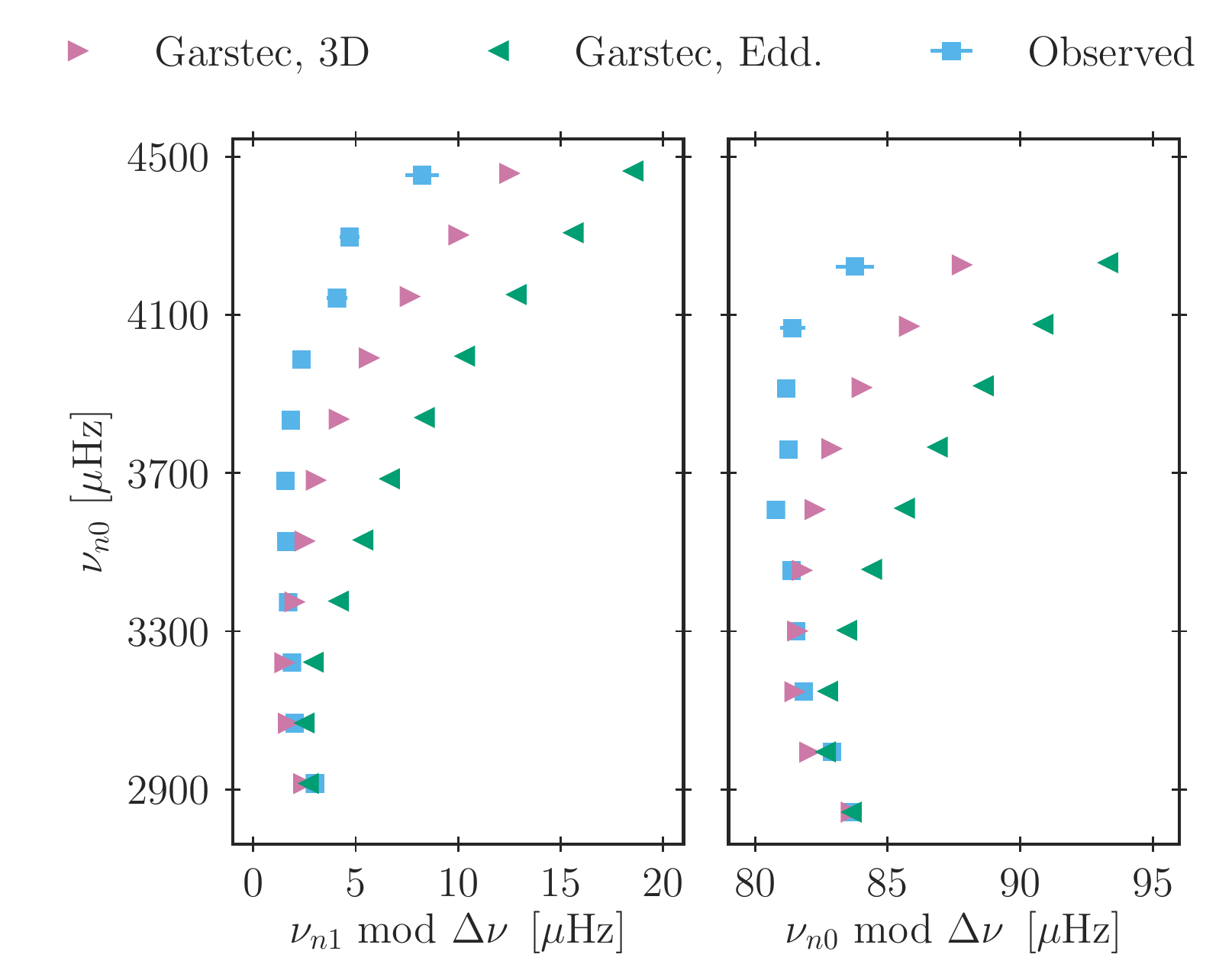}
  \caption{\'{E}chelle diagrams of KIC~9955598 for $\ell=0$ (right) and $\ell=1$
    (left). Model frequencies are from the best-fitting model obtained with
    \bas, using the two different grids of stellar models with Eddington
    atmospheres and \tdes \otf, respectively (details in the text)}
  \label{fig:KIC9955598_echelle}
\end{figure}

Moving on to the other case, KIC~9955598, we return to the surface effect and
the stellar evolution models. The frequencies of the two BFM's -- obtained from
the grid-based fit with an additional constraint on the lowest $\ell=0$ mode --
and the observations of KIC~9955598 can been seen in
\fref{fig:KIC9955598_echelle}. The comparision is shown in the form of
\'{e}chelle diagrams for two different radial orders. From the figure it is very
clear that the model appending \tdes \otf has frequencies deviating
significantly less from the observed values for both orders. All in all,
compared to canonical stellar evolution models, the oscillation frequencies from
our new models are much closer to the observations, without the use of any sort
of correction for the surface term.

\section{Red-giant branch models}
\label{sec:rgb}

In line with the previous section, we investigate further the asteroseismic
implications of our new method, focusing here on the solar-like acoustic
oscillations in red-giant stars and how the surface effect changes. Red giants
are very important for many astrophysical fields, e.g. for probing distant
regions in the Galaxy or studying star clusters. The analysis of such stars has
matured rapidly in the era of space-based photometry and asteroseismology with
\corot \citep{Baglin2009} and \kep \citep{Borucki2010x,gilliland10a}, but the
modelling deficiencies in the near surface layers are still not well understood.

In order to perform a detailed differential frequency analysis of two stellar
models, the two must be very similar seismically. It is natural to compare
models of identical mean density, which is correlated with the asteroseismic
large frequency separation \dnu \citep{ulrich86a}. To ensure this, we adopted
the convergence criterion devised for the \textsl{Aarhus Red Giants
  Challenge}\footnote{A series of workshops dedicated to modelling of red-giant
  stars and especially to detailed comparisons of many different stellar
  evolution codes.} (\papRGp; \papRGiip). For a given model (\textsl{mod}) the
minimum acceptable convergence at $a$ solar masses and $b$ solar radii is
defined as
\begin{equation}
  \Delta_{\mathrm{convergence}} = \left| 1 - \dfrac{G_{\mathrm{mod}}M_{\mathrm{mod}}/R_{\mathrm{mod}}^3}%
    {G (a \times \msun)/(b \times \rsun)^3} \right| \leq 2 \times 10^{-4} \; ,
\end{equation}
where $G$ is the gravitational constant. The choice of $2 \times 10^{-4}$ is a
compromise between the uncertainties in the asteroseismic frequencies and the
ease of finding the required model, as discussed by \papRGt.

For models calculated with the same stellar evolution code, $G$ is of course
invariant and the convergence is solely determined by the mass and radius. For
the model to match a reference model with the desired radius $R_{\mathrm{ref}}$,
the criterion can be rewritten and reduced to
\begin{equation}
  \label{eq:rg_match}
  \Delta_{\mathrm{convergence}} = \left| 1 - \dfrac{M_{\mathrm{mod}} R_{\mathrm{ref}}^3}%
    {M_{\mathrm{ref}} R_{\mathrm{mod}}^3} \right|
  \leq 2 \times 10^{-4} \; ,
\end{equation}
where $M_{\mathrm{ref}}$ is the mass of the reference model at radius
$R_{\mathrm{ref}}$.

For this analysis we use the same settings in the computations as in
\sref{sec:evol} and \ref{sec:kepler} -- i.e. like those listed in
\sref{sec:method}, but without microscopic diffusion -- and two initial masses:
$1.00$ and $1.30 \msun$. Like in the previous section, two different sets of
models were calculated: One appending Stagger \tdes \otf and another using a
standard Eddington \ttaurel. The models with Stagger \tdes are taken as the
reference model in \eref{eq:rg_match} and the Eddington model is carefully
calculated to match within the convergence limit. For the comparison we selected
RGB models at different surface gravity positions -- $\logg = 3.0$, $\logg =
2.5$, and $\logg = 2.0$ -- and an overview can be seen in \tref{tab:rgb}
(alongside the results described below).
\begin{table*}
	\centering
	\caption{Comparison points for RGB models. The convergence in radius is
    according to \eref{eq:rg_match} with Stagger \tdes \otf as the reference,
    and Eddington computed to match. The frequency of maximum oscillation power,
    \numax, is determined from the Stagger models using \eref{eq:numax}. The
    frequency difference is \enquote{3D $-$ Eddington} and determined as a
    3-point average around \numax (see the text for details).}
	\label{tab:rgb}
  \begin{tabular}{rrrrrr}
    \hline
    $M$ [\msun{}] & $\log g$ & $R$ [\rsun{}] & \numax \muhz & $\delta\nu_{\numax}$ \muhz & $\delta\nu_{\numax} / \numax$ \\
    \hline
    $1.00$        & $3.0$                 & $5.24$        & $124.49$           & $-0.510$                         & $-4.10 \times 10^{-3}$         \\
    $1.00$        & $2.5$                 & $9.32$        & $40.34$            & $-0.309$                         & $-7.67 \times 10^{-3}$         \\
    $1.00$        & $2.0$                 & $16.47$       & $13.31$            & $-0.097$                         & $-7.31 \times 10^{-3}$         \\[3pt]
    $1.30$        & $3.0$                 & $5.97$        & $123.75$           & $-0.564$                         & $-4.56 \times 10^{-3}$         \\
    $1.30$        & $2.5$                 & $10.63$       & $40.10$            & $-0.228$                         & $-5.70 \times 10^{-3}$         \\
    $1.30$        & $2.0$                 & $18.89$       & $13.11$            & $-0.104$                         & $-7.95 \times 10^{-3}$         \\
    \hline
\end{tabular}
\end{table*}

\begin{figure}
  \centering
  \includegraphics[width=\linewidth]{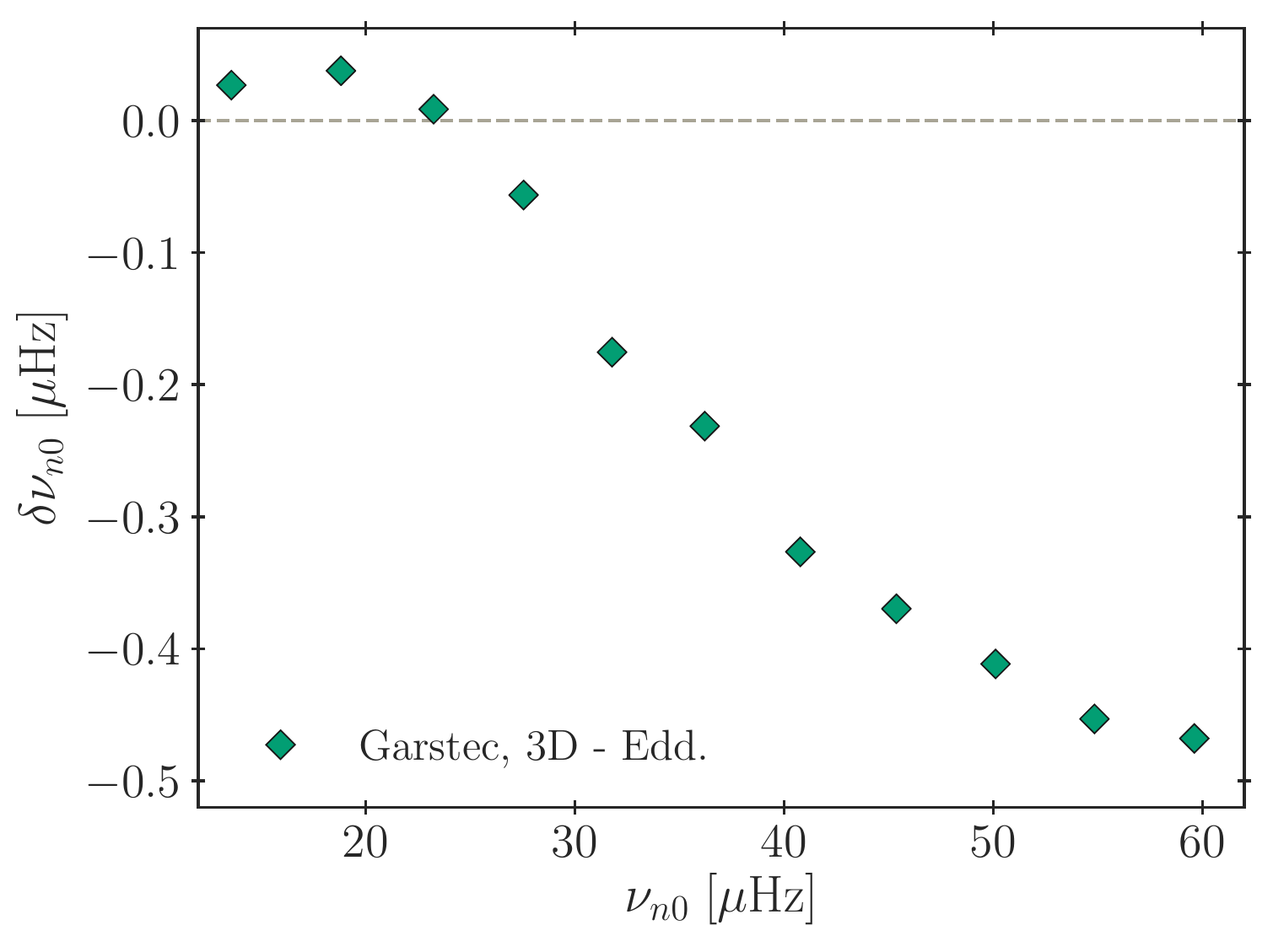}
  \caption{Frequency difference between calculated $\ell = 0$ frequencies for
    the model with \tdes appended \otf and standard Eddington model with initial
    masses of $1.00 \msun$, as a function of \tde frequencies. The comparison
    point is chosen near $\logg = 2.5$ at $R=9.32 \rsun$.}
  \label{fig:rgb-freqdiff}
\end{figure}

For a set of matching models, the oscillation frequencies are computed with \adi
and compared. An example of the frequency difference comparison for models with
$M = 1.00 \, \msun$ at $R = 9.32 \, \rsun$ is shown in \fref{fig:rgb-freqdiff}.
For the other comparison points, the shape of the differences looks almost
identical, albeit with lower frequencies and fewer modes the lower \logg gets.

To quantify this, and to compare the actual shape of the deviation between the
different comparison points, we need to scale the quantities. Thus, we calculate
the frequency of maximum oscillation power, \numax, using the scaling relation
from \citet{kjeldsen95a},
\begin{equation}
  \label{eq:numax}
  \frac{\numax}{\nu_{\mathrm{max} , \odot}} \simeq \left(\frac{M}{\msun}\right) \left(\frac{R}{\rsun}\right)^2 \left(\frac{\teff}{\tsun}\right)^{-1/2} \; ,
\end{equation}
where $\odot$ denotes the solar values, and specifically
$\nu_{\mathrm{max},\odot} = 3090 \, \muhz$. For a given set of converged models,
\numax is calculated using the quantities of the \tde model -- due to the
convergence defined by \eref{eq:rg_match}, \numax of the two models in a pair
are almost identical, with the variation caused by the differences in \teff (on
the order of $40\kel$). Then we select the oscillation mode closest to \numax
and the two adjacent modes -- one on either side -- and calculate the average
frequency deviation of these three modes, which we denote $\delta\nu_{\numax}$.
Now, the frequencies are scaled by \numax and the frequency differences by
$\delta\nu_{\numax}$.

\begin{figure}
  \centering
  \includegraphics[width=\linewidth]{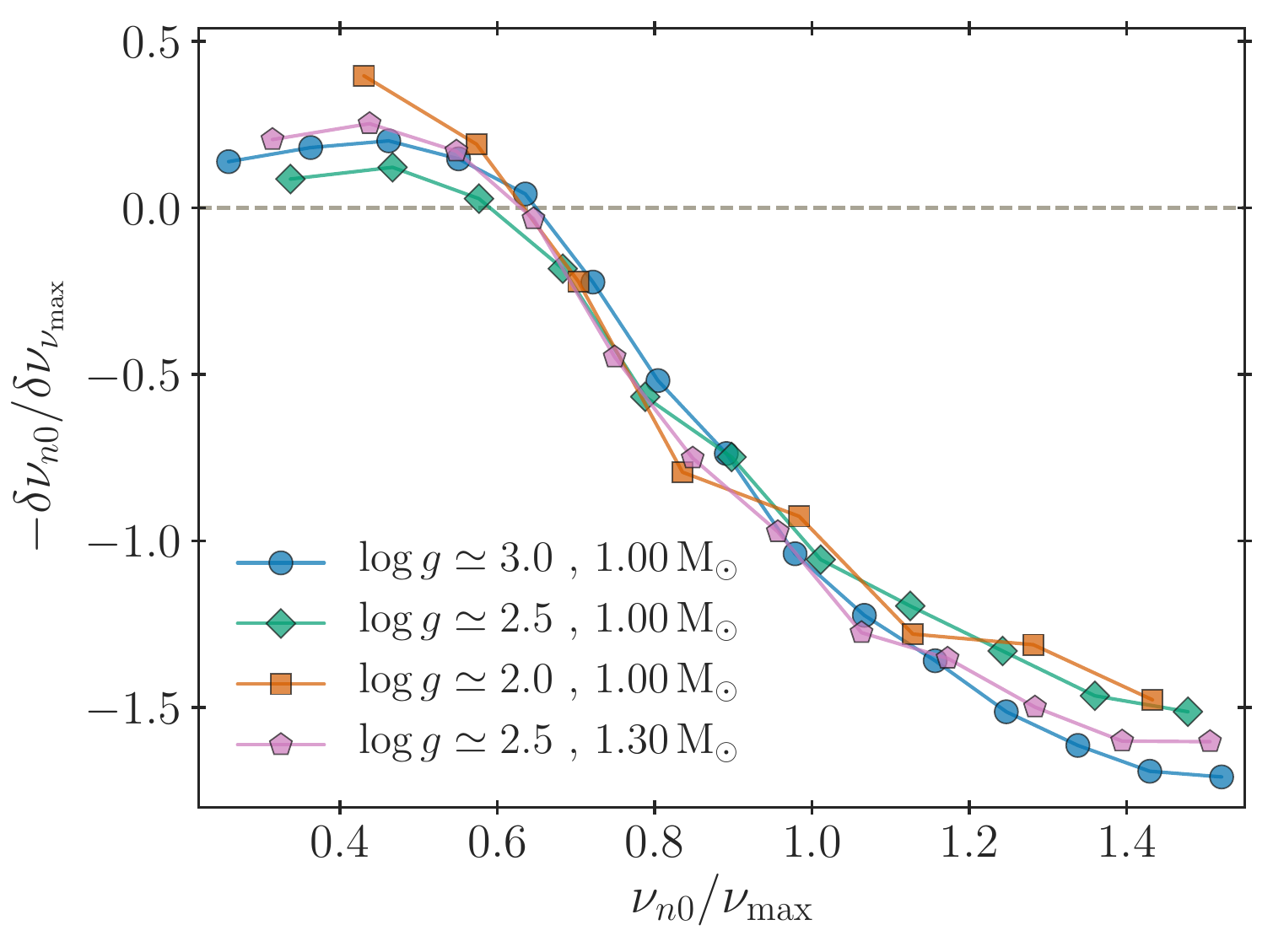}
  \caption{Scaled frequency differences (\enquote{3D $-$ Eddington}) for several
    RGB models. For a given comparison, the model frequencies and frequency
    differences are scaled by respectively the frequency of maximum oscillation
    power, \numax, and the difference at this point, $\delta\nu_{\numax}$
    (details in the text) The green diamonds are identical to those in
    \fref{fig:rgb-freqdiff}, except for the applied scaling. All quantities and
    comparison points are listed in \tref{tab:rgb}}
  \label{fig:rgb-scaled-freqdiff}
\end{figure}

All of the quantities are listed in \tref{tab:rgb}, and in all cases the
relative difference at the frequency of maximum power is below $1 \%$.
Additionally, four of the resulting curves -- which turns out to be remarkably
similar -- are shown in \fref{fig:rgb-scaled-freqdiff}. From this figure, it is
clear that the shape of the deviation is independent of position on the RGB, and
equally important independent of mass as well.

As also mentioned in the introduction, the application of atmospheric 3D
hydro-simulations to study the surface effect of solar-like oscillators has been
the focus of different studies in the recent years. Several groups have utilized
the patched model technique to perform investigations across the HR (or Kiel)
diagram. One to highlight in the current context is the work by
\citet{Sonoi2015}, because two of their comparison points are denoted as red
giants: Their model I ($\teff = 5885\kel$, $\logg=3.5$) and model J
($\teff=4969\kel$, $\logg=2.5$). Even though it is somewhat hotter, the latter
is especially interesting for our comparison being furthest up the RGB. Note,
however, that the performed patch for J is using a $M=3.76\msun$ model, which is
very different from our cases.) They derive corrections both in the form of the
classical \citet{Kjeldsen2008} power law and their own \enquote{modified
  Lorentzian formulation}, where one of the quantities in both fits can be
directly translated to $\delta\nu/\numax$ at $\nu = \numax$ (denoted
$\delta\nu_{\numax}/\numax$ above). We predict the same sign of the deviation
and, taking the quite different approaches into account, our results are roughly
of the same magnitude. Furthermore, their equation 10 provides the fitting
factors as a function of \logg and \teff; we do not see a clear surface gravity
trend as they predict, but the magnitude of the estimates from this is also in
line with our findings.

However, as very recently shown by \cite{Joergensen2019} based on an analysis of
315 patched models, the coefficients in the Lorentzian formulation by
\citet{Sonoi2015} strongly depend on the underlying sample. Consequently, there
exists no set of coefficient values that is universally applicable throughout
the parameter space. The authors note that the formulation derived by
\citet{Sonoi2015} suffers from a selection bias, as it is is predominantly based
on models, for which $\teff > 6000 \kel$ and $\logg \geq 4.0$. It follows that
their fit cannot be directly applied to stars on the RGB. Our current results
are in line with this conclusion.

Finally, we note that neither the Lorentzian formulation by \citet{Sonoi2015}
nor the frequency difference we present in \fref{fig:rgb-scaled-freqdiff} can be
directly translated into a surface correction relation, since modal effects and
turbulent pressure have been neglected.

\section{Conclusions}
\label{sec:conclusions}

We have presented an extensive analysis of stellar models that append \tdes \otf
from the \staggrid \citep{Magic2013} at each time step, following the procedure
introduced by \papI. These models provide a more physically accurate description
of the outermost layers in low-mass stars.

When calculating the appended \tdes, we use the equation of state from the
stellar evolution code. In \papI, we verified that the density and temperature
obtained from the equation of state showed good agreement with the 3D solar
simulation. Here we verified that the resulting first adiabatic index \gam does
not significantly shift the obtained oscillation frequencies compared to using
\gam directly from the 3D simulation.

By performing different solar calibrations, we investigated the effect of the
so-called matching point, i.e., the depth above which the \tdes are appended to
the stellar model. We find the mixing length to increase monotonically with
increasing matching depth, which is in qualitative agreement with
\citet{Schlattl1997}. This being said, the evolutionary tracks are relatively
insensitive to the matching point, provided that it is placed sufficiently deep
within the superadiabatic outer layers. Moreover, we find that the oscillation
frequencies are equally independent of the matching depth for sufficiently high
values.

We have performed a solar model analysis using the grid of 3D simulations
computed by \citet{Trampedach2013}, and find consistency with the \staggrid
results (shown in \papI). Moreover, the computed evolutionary tracks are shifted
towards higher effective temperatures on the red-giant branch (RGB) compared to
reference Eddington-grey models. The same qualitative effect was found by
\citet{Mosumgaard2018} utilising parametrised information from the same 3D grid
extracted by \citet{Trampedach2014a,Trampedach2014b}.

Moving on from the Sun, we have computed evolutionary tracks for stars of
different mass to further test our procedure for including \tde \otf. The tracks
show prominent kinks at the boundaries of the grid of 3D simulations, as well as
in regions where the sampling is sparse -- especially problematic on the RGB and
in the PMS. This calls for an refinement of the 3D grids to make the
interpolation more reliable (see also \pp). Furthermore, more simulations at
higher temperatures will extend the usefulness of our method by widening the
allowed mass range.

Moreover, we took advantage of the different models across the $(\teff,
\logg)$-plane to further investigate the applicability of our method and
specifically the equation of state. By comparing to full 3D simulation, we can
conclude that the density is reproduced accurately for main sequence stars and
subgiants; however the residuals are slightly larger for more evolved giants.
This is partially a resolution effect, due to the very few simulations along the
RGB, where the tracks are close to each other.

For the first time, an asteroseismic analysis using stellar models including
\tde \otf is presented. Using a grid-based approach and the Bayesian inference
code \bas \citep{Aguirre2015}, we determined the stellar parameters of two stars
from the \kep \leg sample \citep{Lund2017,Aguirre2017}. We find that the
obtained parameters are consistent -- in the sense that they agree within the
uncertainties -- between the grid with our new method and the reference
Eddington case. This consistency also holds between fits to individual
frequencies and frequency separation ratios. Furthermore, comparing the
best-fitting models from both grids to the observations, we see that the
asteroseismic surface effect is strongly reduced by using \tdes \otf. In other
words, our new models are able to predict frequencies much closer to
observations without using any additional corrections -- and are able to do so
consistently across stellar parameters.

Finally, we extended the asteroseismic investigation to red giants. We carefully
matched standard Eddington models with \tdes appended \otf to look at the
detailed differences in their oscillation frequencies. We see a relative
difference below $1 \%$ at the frequency of maximum power, and no trend in shape
or relative deviation with either mass or surface gravity.

Although the differences between the new models and the patched ones -- and even
the Eddington-grey ones (besides the surface term) -- are rather minor, we have
demonstrated the robustness of the method with regard to details of the
application of the \tdes, both for main-sequence and red-giant stars. We expect
larger effects for lower metallicity red giants, which we treat in a future
work. As a last remark, we want to once again stress the need for a denser grid
of 3D atmospheres/envelopes, and the requirement to compute new hydrodynamical
simulations to achieve this. This will enable the use of \tdes \otf for stellar
evolution -- to obtain more realistic models -- to reach its full potential.

\section*{Acknowledgements}

We thank the anonymous referee for the thorough review and useful suggestions,
which have improved the presentation of the paper. We give our thanks to R.
Trampedach for many insightful discussions, his opinion on technical matters and
for providing us with his grid of simulations. We also record our gratitude to
R. Collet, Z. Magic and H. Schlattl for a fruitful collaboration. We acknowledge
the funding we received from the Max-Planck Society, which is much appreciated.
Funding for the Stellar Astrophysics Centre is provided by the Danish National
Research Foundation (grant DNRF106). ACSJ acknowledges the IMPRS on Astrophysics
at the Ludwig-Maximilians University. VSA acknowledges support from VILLUM
FONDEN (research grant 10118) and the Independent Research Fund Denmark
(Research grant 7027-00096B). This research was partially conducted during the
Exostar19 program at the Kavli Institute for Theoretical Physics at UC Santa
Barbara, which was supported in part by the National Science Foundation under
Grant No. NSF PHY-1748958




\bibliographystyle{mnras}
\bibliography{manual_refs,mendeley_export}


\appendix

\section{Grid morphology}
\label{app:morph}

The \staggrid is designed with spectroscopy in mind and is therefore regular in
the Kiel diagram, i.e., in the (\teff, \logg)-plane. However, as described in
the paper, we also need to set up a triangulation in the (\tma, \logg)-space,
where \tma is the temperature at the matching point given the choice of \km.
This is required by our implementation in order to infer \teff by interpolation.

In \sref{sec:sun_depth} and \sref{sec:evol_tracks}, it was discussed that the
effective resolution of \staggrid in the (\tma, \logg)-plane is different from
the almost-regular (\teff, \logg)-space. This is an effect of the given
simulation points moving as a function of \km.

\begin{figure}
  \centering
  \includegraphics[width=\linewidth]{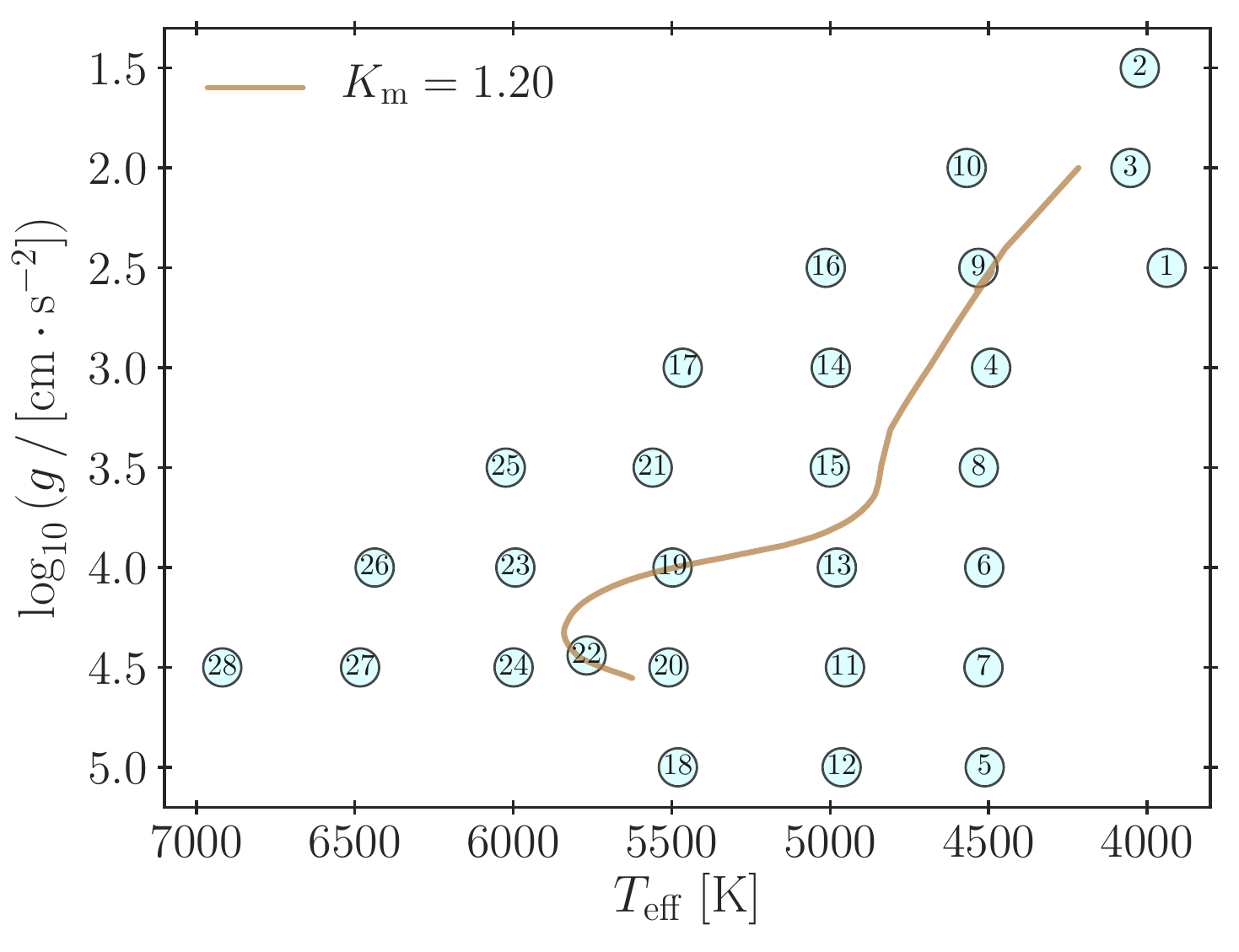}
  \includegraphics[width=\linewidth]{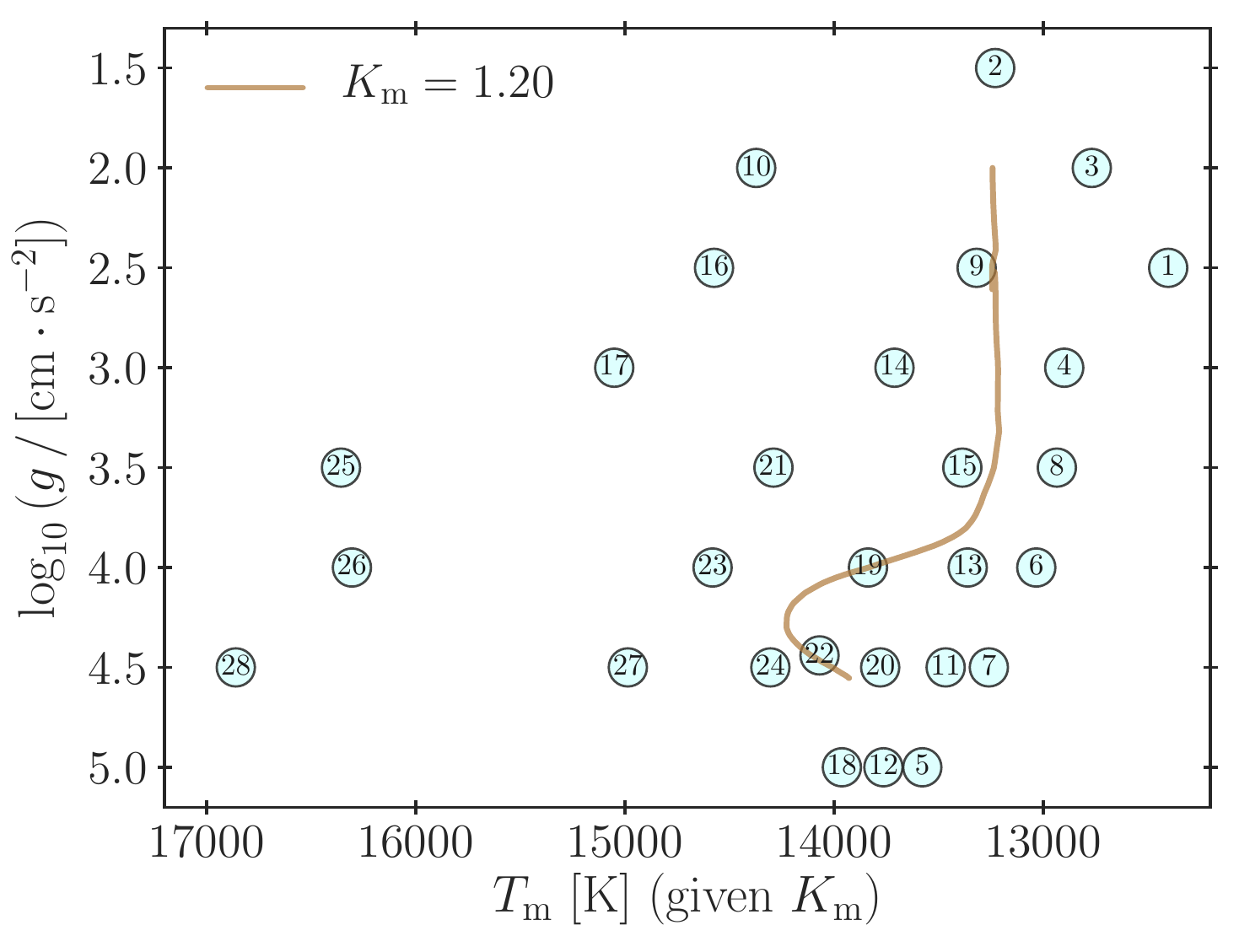}
  \caption{The \staggrid at solar metallicity. The simulation points are
    numbered according to \teff. The evolutionary track is for a $1.00\msun$
    star using \tde \otf. \emph{Top panel:} Kiel diagram with the \staggrid at
    solar metallicity. \emph{Bottom panel:} Temperature at the matching point,
    \tma, for $\km=1.20$.}
  \label{fig:app-grid-kiel}
\end{figure}

In the top panel of \fref{fig:app-grid-kiel}, the \staggrid at solar metallicity
is shown in the traditional Kiel diagram. To ease the discussion, the individual
simulation is annotated with a number -- $1$ for the lowest \teff and $28$ for
the highest. In the bottom panel of the figure, the grid is shown in the form of
\tma and \logg for our usual choice of $\km = 1.20$. The simulation points in
this figure have the same labels as in the first one. In both figures, a $1.00
\msun$ track using our new implementation with $\km = 1.20$ is shown for
reference.

From the plots, it is clear that the morphology of the grid changes. It is also
evident that the \enquote{movement} of the points depends on the surface
parameters of the simulation. This has the profound effect that some of the
simulation points switch places in the temperature ordering. As an example,
simulation $2$ is colder than $3$ looking at \teff, but hotter in \tma; the
opposite is the case for $8$ and $6$, where the former has the highest \teff,
but the latter highest \tma. Another trend is the tendency for the cold
simulations to \enquote{clump}, which is very evident looking at $[5,~12,~18]$
or $[7,~11]$. On the other hand, simulations like $[3,~10]$ or $[9,~16]$ instead
move away from each other.

These changes in the temperature ordering can affect the triangulation, thus
altering the interpolation results; thus, this is directly linked to the
observed kinks in the evolutionary tracks in the Kiel diagram.

\bsp	
\label{lastpage}
\end{document}